\documentclass[options]{mn2e}

\title[Kinematically quiet halos around distant radio galaxies]
{Kinematically quiet halos  around $z\sim$2.5 radio galaxies. 
Keck spectroscopy\thanks{Based on: a) observations made at the W.M. Keck Observatory which is operated
as a scientific partnership among the California Institute of Technology,
the University of California and the National Aeronautics and Space Administration
(NASA). The observatory was made possible by the generous financial support
of the W.M. Keck Foundation 
b) observations with VLT Antu unit   at the European 
Southern Observatory, Paranal, Chile  
c) observations with the NASA/ESA Hubble Space
Telescope, obtained at the Space Telescope Science Institute, which is operated by
the association of Universities for Research in Astronomy
 d) observations made with the NRAO Very Large Array.
The National Radio Astronomy Observatory is a facility of the National
Science Foundation operated under cooperative agreement by Associated Universities
 Inc.}}
\author[Villar-Mart\'\i n et al.]{M. Villar-Mart\'\i n$^1$\thanks{At the 
Instituto de Astrof\'\i sica de Andaluc\'\i a (Granada, Spain) from Sept 1st 2003}, J. Vernet$^2$,  
S. di Serego Alighieri$^2$, R. Fosbury$^3$,
\newauthor  A. Humphrey$^1$, L. Pentericci $^4$ \\
$^1$Dept. of Physical Sciences, University of Hertfordshire, College Lane, Hatfield,
Herts, AL10 9AB, UK\\
$^2$Osservatorio Astrofisico di Arcetri, Largo E. Fermi 5, I-50125, Firenze,
 Italy\\
$^3$Space Telecope European Coordinating Facitily,
Karl Schwarschild Str. 2, D-85748 Garching bei Muenchen, Germany\\
$^4$Max Plank Institute fur Astronomie, Konigstuhl 17, D-69117 Heidelberg, 
Germany\\}

\date{}

\begin{document}

\pagerange{\pageref{firstpage}--\pageref{lastpage}} \pubyear{2002}

\maketitle

\label{firstpage}

\begin{abstract}
     We present the results of the kinematic study of the 
extended gas
in a sample of 10 high redshift radio galaxies ($z\sim$2.5)
 based on  high signal to noise Keck II and VLT long slit
spectroscopy. In addition to the typical  high
surface 
brightness kinematically perturbed regions (FWHM and velocity shifts
$>$1000 km s$^{-1}$), we find in all objects
giant low surface brightness  halos  which show 
quieter kinematics with typical emission line FWHM and velocity
shifts of $\sim$several hundred km s$^{-1}$.

The giant halos often extend for more than 100 kpc and sometimes beyond 
the radio 
structures.  They emit lines 
other than Ly$\alpha$ (CIV, HeII and NV in some cases), 
typically found in the spectra of high redshift active galaxies.
 Continuum is also often detected. 
The halos are enriched with heavy elements at tens of kpc from the 
active nucleus. Typical Ly$\alpha$ luminosities and surface 
brightness  (within the slit)  are in the range 10$^{43-44}$ erg s$^{-1}$
and several $\times$ 10$^{-17~to~-16}$ erg cm$^{-2}$ s$^{-1}$ arcsec$^{-2}$
 respectively. Estimated densities are in the range
$\sim$17-150 cm$^{-3}$.
The 
  quasar
continuum is the dominant source  of ionization of the quiescent halos  
along the radio
axis. The implied total quasar ionizing
luminosities  are in the range $\sim$several $\times$ 10$^{45}$-10$^{47}$
 erg s$^{-1}$,
in the same range as radio loud quasars at comparable redshift.

The detection of giant quiescent halos in all objects suggests that
they could be a common ingredient of high redshift radio galaxies. 
The radio galaxies seem to be embedded within the halos.
The nature, origin of the halos and the cosmological
implications are also discussed.

\end{abstract}


\section{Introduction}

Our current belief is that the hosts of powerful radio sources in the distant
Universe  are destined to become the giant ellip\-ticals of today, the most
massive  galaxies we know (e.g.  Lilly \& Longair 1984, McLure et al. 1999). 
They inhabit rich
environments that are believed to be galaxy clusters in the process of
formation (e.g. Pentericci et al. 2000,
 Venemans et al. 2002). While some
distant powerful radio sources may have commenced their formation at very high
redshift, the process of assembly is very active at $z\sim$2-3.  This
corres\-ponds to the epoch when luminous quasars appear to have had their maximum
space density and the star formation rate of the Universe was two orders of
magnitude higher than at present (e.g. Pei 1995,
Shaver et al. 1996,
Fan 2001).

 Our group has recently proved that solar or supersolar metallicities are
common  in the extended gas (tens of kpc) of radio galaxies at $z\sim$2.5 
(Vernet et al. 2001, Villar-Mart\'\i n et al. 2001). 
The relatively high metallicities imply that star formation in these objects has
been rapid and intense, ena\-bling a fast chemical enrichment of the gas. 
Since stellar population analysis of giant ellipticals at low redshift 
indicates that most stars have been formed during a brief epoch with high star
formation rate, this  further supports the idea that 
powerful high redshift  radio galaxies  are progenitors of the giant 
ellipticals of today. 
Star formation is also suggested by the detection at submm wavelengths 
of some high redshift radio galaxies
(e.g. Archibald et al. 2001, Reuland et al. 2003a).
 Although the nature of the heating mechanism
is controversial, some authors have argued that 
 the dust is heated by  starbursts, rather than the hidden
quasar nuclei (e.g. Tadhunter et al. 2002).
By looking at high redshift
radio galaxies ($z\ga$2-3, hereafter HzRG) we are therefore likely to be
witnessing the formation process of giant elliptical galaxies.

	HzRG are surrounded by giant nebulae which often extend
for more than one hundred kpc (e.g. McCarthy et al. 1990a, Reuland et al. 2003b)
 and sometimes beyond the radio structures
(e.g. Eales et al. 1993, Kurk et al. 2001, Maxfield et al. 2002). 
 	One of the great difficulties in the interpretation of the optical
and near infrared data  of HzRG is due to 
 the fact that the observed morphological, kinematic
and ioni\-zation properties are strongly influenced  by the nuclear activity
and the extreme phenomena associated with the rela\-tivistic jets.
 Most morphological and kinematic 
stu\-dies of HzRG have been focussed on the high surface brightness regions. 
These are clumpy, irregular (with features such as fila\-ments, plumes,
ionization cones, e.g. Reuland et al. 2003b)  and often aligned with the 
radio axis (Chambers, Miley \& van Breugel 1987, McCarthy et al. 1987).
They are characterized by extreme kinematics, with measured 
FWHM and velo\-city shifts  $\geq$1000 km s$^{-1}$ 
(e.g. McCarthy, Baum \& Spinrad 1996, Villar-Mart\'\i n, Binette \& Fosbury 1999), compared to values
of a few hundred in low redshift radio galaxies (e.g. Tadhunter, Fosbury \& Quinn
1989, Baum, Heckman \& van Breugel 1990). 
It is likely that a perturbing mechanism such as shocks 
induced during jet/gas interactions \cite{ojik97} 
 is responsible for
such extreme kinematics. 

	However, there
is evidence in two HzRG for
giant (distance to the nucleus $\ga$ 50 kpc\footnote{We assume
$\Omega_{\Lambda}$=0.7, $\Omega_M$=0.3, H$_0$=65 km s$^{-1}$ Mpc$^{-1}$})
 low surface brightness  
halos (LSBHs hereafter) of 
ionized gas with {\it quiet  kinematics}:
1243+036 at $z=$3.6 (van Ojik et al. 1996, vO96 hereafter) 
and 0828+193 at  $z=$2.6 (Villar-Mart\'\i n et al. 2002, VM02 hereafter).

These  quiescent  halos might be residual gas from which
the galaxy started to form (VM02). They offer opportunities to study
 the ionizing photon luminosity of the quasar 
 and with sufficient line measurements the 
chemical composition of the halo gas. 
If the gas kinematics is gravitational
in origin the halos can
be used as mass tracers of radio galaxies at different redshifts. 
They   would provide
 critical information to constrain models of galaxy formation and evolution. 
Also, the metallicities of the giant halos will set tight constraints
on their chemical enrichment history and, ultimately, on the formation process
of the  galaxy.

 We present in this paper the results of  
a research project whose goal is to search for 
kinematically unperturbed  halos in HzRG and study their kinematic and
ionization properties. 	An important advantage of the current 
set of data is that we can use HeII (and CIV
in some cases) to constrain the kinematic properties of the gas, rather than using
only Ly$\alpha$, which is very sensitive to dust/gas absorption.

	The paper is organized as follows: The observations and data
analysis are described in \S2. General results and results on individual
objects are presented in \S3.  The ionizing mechanism, kinematic 
patterns, nature of the 
giant quiescent halos and cosmological implications
are discussed in \S4.  The summary and conclusions are presented in \S5.

\section{Observations and Data Analysis}

\subsection{The data}

All the sources were selected from the ultra-steep spectrum 
(USS) radio galaxy survey (see e.g. 
R\"ottgering et al. 1994) with redshift $z>$2.2.

	The spectra for all the objects (see Table 1) except 2104-242
were obtained  with the Low Resolution Imaging Spectrometer 
(LRIS, Oke et al. 1995) with its polarimeter (Goodrich, Cohen, Putney  1995) 
at the Keck II 10 m teles\-cope from July 1997 till July 1999 
under subarcsec seeing conditions (seeing varying between $\sim$0.7 
and 1.1 arcsec).
The observing log is presented in Vernet et al. (2001).

The LRIS detector is a TEK 2048$^2$ CCD with 24 $\mu$m pixels giving a scale
of 0.214 arcsec pixel$^{-1}$. We
used a 300 line mm$^{-1}$ grating  which provides 
a dispersion of 2.4 \AA ~  pixel$^{-1}$ and a spectral range $\lambda_{obs}
\sim$3900--9000 \AA . The slit width was always 1 arcsec, except for 4C-00.54 and 
4C+23.56 for which we used a 1.5 arcsec slit. 	The slit was always oriented along the radio axis
given by R\"ottgering et al. 1994 and Carilli et al. 1997 (see Table 1).

 Each individual frame was calibrated in
wavelength and corrected for slit curvature using arc spectra. This
initial wavelength calibration was refined using strong sky lines.
The FWHM of the instrumental profile (IP, Table 1) was determined 
from unblended sky lines. These  values corres\-pond
 to a range in resolution of
320--610 km s$^{-1}$ at the redshifted HeII$\lambda$1640 wavelength. Variations from object to object 
 are probably due to mechanical instabilities within the
instrument causing variation on the line spread function.

In order to compensate for the effects of flexure in the instrument,
 we extracted from every frame a section around Ly$\alpha$ and NV$\lambda$1240 and one
around CIV$\lambda$1550 and HeII$\lambda$1640. 
The individual sub-frames corresponding to the
same  spectral range were spatially 
aligned (using the con\-tinuum centroid) and combined. The two final frames (one for
Ly$\alpha$ and NV  and another one for CIV and HeII) were also spatially aligned, assuming
that the continuum centroid has the same spatial position at all wavelengths. This is
reasonable in this small spectral range, taking also into account that the spatial
resolution element has a large physical size ($\sim$ 8-9 kpc). 
For a more detailed description of the observations  and data reduction 
see Vernet et al. (2001).

	The observations for 2104-242
 were carried out in service mode on 1999 September 2-5 with 
FORS1 on the 8.2 m VLT Antu telescope (ESO-Chile).
The 600B grism was used with a 1 arcsec wide slit. The slit was positioned 
along the brightest components and the filamentary structure at a 
position angle of 2$\degr$ north
through east. The exposure time was 10800 sec.
 The seeing during the observations was 1  arcsec and conditions were photometric. 
A 2 $\times$ 2 readout binning was used in order to increase
 the signal-to-noise ratio (S/N). The resultant spectral
resolution was 6 \AA\ (FWHM) and the spatial scale was 0.4 arcsec pixel$^{-1}$. 
For a detailed description of the data reduction
see Overzier et al. (2001).

	Radio maps, optical and near infrared HST archive ima\-ges were also used 
and registered with the Keck 2D spectra for comparison between the structures
observed at different wavelengths. 
	The radio observations and data are described by Carilli et al. (1997)
and the HST WFPC2  and NICMOS images are described in Pentericci et al. (1999
and 2001) respectively.

\begin{table*}
\centering
\begin{tabular}{lllllll}
\hline
  IAU  &  4C 		&  Redshift & Exp.  & PA  &  IP & HST \\ 
    & 		&  & sec  & $\degr$ &  \AA  &  images\\ \hline
1809+407    &   4C+40.36     	& 2.265	& 10742 & 82   & 9.8$\pm$1.2 & No \\ 
0211-122    &    		& 2.340	& 28580 & 104  &  9.5$\pm$1.2  & WFPC2/NICMOS \\ 
1931+480    &   4C+48.48   	& 2.343	& 12000 & 50   & 11.2$\pm$1.2 & No \\ 
1410-001    &  4C-00.54     	& 2.360	& 16240 & 134  & 14.3$\pm$1.2  &  WFPC2/NICMOS \\ 
0731+438    & 4C+43.15   	& 2.429	& 22720 & 12  &  10.5$\pm$1.2  & No \\ 
2105+236    &   4C+23.56    	& 2.479	& 22266 & 47   & 13.6$\pm$1.2 & WFPC2 \\ 
2104-242    &       		& 2.491	& 10800 & 2  & 6.0$\pm$0.5 &  WFPC2/NICMOS \\ 
1558-003    &         		& 2.527	& 11400 & 72  & 11.0$\pm$1.2 & No \\ 
0828+193   &			& 2.572 & 18000 & 44 & 10.5$\pm$1.2  &  WFPC2/NICMOS \\
0943-242    &     		& 2.922	& 13800 & 73   & 10.8$\pm$1.2 & WFPC2 \\ 
\hline
\end{tabular}
\caption{The radio galaxy sample. Objects have been organized in order of
increasing redshift. The exposure times, 
slit position angles (PA) and instrumental
profiles FWHM (IP) are also shown. Results for 0828+193 have been 
presented in Villar-Mart\'\i n et al. 2002. The last column shows the HST images used in the
regitration process between the radio/HST images and  Keck spectra (\S2.2)}
\end{table*}

\subsection{Registration}

Several methods of registration between the radio maps and the 2D  
spectra were used, depending on the availability of HST images and on  
the presence of a radio core.

For objects presenting a radio core and for which we have both WFPC2 and
NICMOS images (0211-122, 0828+193, 1410-001 and 2104-242), the IR peak was
assu\-med to coincide with the radio core. The WFPC2 image was then registered
with the NICMOS image using field stars. Finally the 2D
spectrum continuum centroid was aligned with the WFPC2 image peak.
The total uncertainty with this method is $\sim$0.3 arcsec. The error
might be larger:  since the AGN is  expected to be obscured,
the near infrared peak probably does
not mark its position and therefore,
it  might be spatially shifted relative to the radio core.

 2105+236 is the only object presenting a radio core for which we only have a
 WFPC2 image. The obvious thing to do would have been to align the
 radio core with the peak of the WFPC2 image. However, the available
 astrometry places the radio core on top of a secondary peak of the WFPC2
 image, close to the Ly$\alpha$ emission vertex in Chambers et al. \shortcite{cham96}
 ground-based narrow band image. We used this  re\-gistration point instead.

Three objects (0731+438, 1558-003 and 1931+480) have a radio core but
 HST images are not available. We simply aligned the radio core with the
peak of the continuum in our 2D spectra. 
The total uncertainty with this method is $\sim$0.2 arcsec.
As above, this error is likely to be larger since the continuum centroid
might not mark the position of the active nucleus.
(Notice that for 1931+480 we have used a different radio core identification
than Chambers et al. 1996).

Two objects do not have a radio core. For 1809+407  the radio 
map and optical spectrum were aligned
using the astrometry from Chambers et al. \shortcite{cham96}. 
We assumed that the optical continuum centroid in the
Keck spectrum coincides with the optical identification by
Chambers et al. The uncertainty of this method is
$\sim$0.5 arcsec.
For 0943-242, we   determined the
astrometry of field stars in the WFPC2 ima\-ge  using the
GSC2.2 catalogue (accuracy of $\sim$0.2-0.3 arcsec)
and then  aligned it with the radio map. 
The continuum centroid  in the 2D spectrum was then aligned 
with the peak of  the image. The uncertainty of this method is
$\sim$0.4-0.5
arcsec.

\subsection{Data analysis method}

	The 2D  spectra 
were divided into several apertures along the spatial
direction and a spectrum was extracted for each one. The apertures were 
selected so that the gas   has  similar 
kinematic properties across the spatial extent  of a given 
aperture and to obtain enough S/N ratio 
in Ly$\alpha$, CIV and HeII
 (if possible) 
to fit
the line profiles. Separate apertures were  defined for the gas beyond the
radio structures when detected.

 The Ly$\alpha$ and HeII spectral profiles were analysed and fitted in 
each aperture
with one or more Gaussians depen\-ding on the quality of the fit. 
 For each kinematic component
the FWHM (corrected for instrumental broadening in quadrature) and the 
velocity shift  relative 
to the HeII line
at the continuum centroid were calculated. HeII has been the 
most important line in our study because it is a 
 strong single (rather than a doublet) 
  non resonant line.
Ly$\alpha$, in spite of its strong sensitivity to absorption by gas and dust  has also often
provided   very valuable kinematic information. CIV is difficult  to constrain due to the
presence of the doublet components. Results for this line are only shown when they provide
convincing evidence for the presence of low velocity gas, a signature of the
quiescent halos.

	Sometimes different fits produced  results of similar quality. 
In most cases it was possible to discriminate by comparing with the
results from  (a) adjacent apertures (b) another  emission
line detected in the same aperture. 
As an example, it was sometimes found that for a given spectrum
one Gaussian component produced as good a fit to the HeII profile as 
two Gaussians components. The profile of Ly$\alpha$ in the same aperture 
and the fits
to the HeII lines in the adjacent apertures provided further
 constraints that sugges\-ted the use of either one or two components.
When such discrimination was difficult, no spectral decomposition
was attempted.

 Any fit which produced lines with
 observed FWHM$_{obs}$ such that

FWHM$_{obs}$  $<$ IP - $|\Delta$IP$|$ -  $|\Delta$FWHM$_{obs}|$

was rejected, where $\Delta$IP is the  uncertainty on the
FWHM of the IP and $\Delta$FWHM$_{obs}$
is the error on FWHM$_{obs}$. A kinematic component was
considered to be unresolved when  FWHM$_{obs}$ was in the range 

IP - $|\Delta$IP$|$ - $|\Delta$FWHM$_{obs}|$ $\leq$ FWHM$_{obs}$
 $\leq$ IP + $ |\Delta$IP$|$ + $|\Delta$FWHM$_{obs}|$.

In such cases an upper limit for the intrinsic FWHM
 was set by estimating the
intrinsic FWHM$_{int}$  for which  
FWHM$_{obs}$ =  IP + $|\Delta$IP$|$ + $|\Delta$FWHM$_{obs}|$.

The formal uncertainty in the central wavelength 
of an emission line is a combination of systematic uncertainties
in the wavelength calibration  (estimated to be $\leq$1 \AA ),
the signal to noise ratio of the line and 
the  errors on  the  Gaussian fits. 
The formal uncertainty in the deconvolved width (FWHM$_{int}$) 
is a combination of uncertainties in the resolution, the 
signal to noise ratio of the 
line and the  errors on  the multiple  Gaussian fits.

The real errors  are likely to be larger  
because the seeing disk might have been smaller than the slit width. 
Although we are confident 
that the seeing was always $\ga$0.7 arcsec, the assumption that the
objects 
filled the slit  introduces additional
uncertainties which will be discussed in \S3.2. Our 
conclusions are not seriously affected by this assumption.

\section{Results}

\begin{table*}
\centering
\begin{tabular}{cccccccll}
\hline
  Name      & $R_{rad}$      &  $R_{max}$ & FWHM    & $V_s$  &  $SB_{Ly\alpha}$  & $L_{Ly\alpha}$  & Detected   \\ \
            &         &  &     &   &   10$^{-17}$ & 10$^{43}$ & Lines   &  \\
 	    &  kpc   &  kpc &  km s$^{-1}$  & km s$^{-1}$ & erg s$^{-1}$ cm$^{-2}$ arcsec${-2}$  & erg s$^{-1}$ 	\\ \hline	
1809+407    &  27     	& 40    & 500-700    	& 400 & 2.0 & 1.6 &   Ly$\alpha$, NV (?), CIV, HeII    \\ 
0211-122    &  62     	& 61    & $\leq$400-700   	& 200 & 1.6 & 1.4 &  Ly$\alpha$, NV, CIV, HeII   \\ 
1931+480    &  110  	&  53   & 450-720  	& 450  & 7.1 & 24.7 &    Ly$\alpha$, CIV, HeII    \\ 
1410-001    &  113     & 62    &$\leq$472-800      & 500  & 2.2 & 10.5 &   Ly$\alpha$,  CIV, HeII \\ 
0731+438    &  42    	&  35   & 600   	& $\leq$100  & 48.9 & 11.6 &   Ly$\alpha$, CIV, HeII     \\ 
2105+236    &  170   &  61   & 500-650  & 570  & 13.6 & 7.4 &  Ly$\alpha$,  HeII      \\
2104-242    &  135    	& 65    & $\leq$265   	& 250 & 2.5  & 0.9 &  Ly$\alpha$, HeII(?) \\ 
1558-003    &  61     	&  69   &  550-850  	& 550 & 2.3 & 13.4 &  Ly$\alpha$, CIV, HeII  \\ 
0828+193    &  64       & 69    &  $\leq$300  	& 600 &  2.4 & 10.2 & Ly$\alpha$, CIV, HeII  \\
0943-242    &  17        & 59    &  400-600   	& 450  &  1.3 &  1.2 & Ly$\alpha$, NV,  CIV, HeII    \\ \hline 
1243+036    &  31        & 78   & 250    	& 450  & & 3.0 &  Ly$\alpha$ (at least)   \\
\hline
\end{tabular}
\caption{Main properties of the quiescent giant halos. 
$R_{max}$ is the maximum extension measured from the position of the continuum
centroid using the Ly$\alpha$ line. $R_{rad}$ is the extension of the radio source measured 
from the position of the
continuum centroid  on  the same side as $R_{max}$.
Column (4) gives  the range of  maximum and minimum FWHM 
values measured across the quiescent halos.
 $V_s$ is the maximum velocity shift measured across the halo.
Both FWHM and $V_s$ refer to the HeII line except for 2104-242 for which
Ly$\alpha$ was used (see text \S3).
 Surface brightness (SB) and luminosities (within the slit) refer to Ly$\alpha$, except
for 0943-242 for which HeII was used (see text \S3). The results
for  1243+036 and 0828+193 were taken from vO96 and VM02 respectively.}
\end{table*}

	Figures 1 to 9 show the data and results of the kinematic study. 
Each figure corresponds to one radio galaxy and
 consists of 3 panels: the top panel  shows the overlay of the radio map
and the WFPC2 image (when available) spatially aligned with the 2D
Ly$\alpha$, NV, CIV and HeII spectra.  The apertures selected
for the kinematic analysis are indicated. The bottom left panel 
 shows the 1D spectra
extracted from  the selected apertures and the bottom right panel shows the
spatial variation of the kinematic properties (velocity shift and FWHM)
of the different 
components revealed by the spectral fits. When two kinematic components 
were isolated, these are distinguished with open  (narrow
components) and solid symbols (broad components). Circles are
used for Ly$\alpha$, triangles  for HeII and squares for CIV. The dashed
vertical lines indicate the maximum extent of the radio structures.

	 The main results of the kinematics analysis are:

	1) All 10 HzRG (including 0828+193, VM02) are associated with
giant halos of quiescent gas   with 
FWHM(HeII)$\leq$850 km s$^{-1}$ and velocity shifts across the nebulae $\leq$600
 km s$^{-1}$.

	2) In addition, 8 out of 10 objects contain kinematically 
perturbed gas. 7 of these objects show a broad component with 
FWHM$>$1500 km s$^{-1}$ and high surface brightness.
2104-242 (which was 
observed at higher resolution) shows signs of kinematic perturbance
in the form of split narrow components. The perturbed gas is usually
located inside the radio structures.

We show in Table 2 some properties of the quiescent LSBHs.  We have included
0828+193 (VM02) and 1243+036 (vO96) for comparison.  
$R_{max}$ is the maximum exten\-sion  measured from the 
position of the continuum
centroid using Ly$\alpha$ (the brightest line). 
Halo sizes are in the range $R_{max}\sim$35-80 kpc\footnote{In the Cosmology we have adopted, 
$R_{max}\sim$78 kpc for the halo detected by vO96
in 1243+036.} and they extend sometimes beyond the radio structures
 (it is possible that radio maps with greater sensitivity might
show that the radio plasma extends further out).
The giant halos emit  lines other than Ly$\alpha$ 
(CIV, HeII and NV in some cases), 
typically found in high redshift active galaxies. The emission 
line spectra  are typical of active
galaxies.
 Continuum is also often detected. In most cases the quiescent gas
is found not only in the outer parts of the objects but  across the high
surface brightness regions as well.

The kinematic parameters  FWHM (maximum and mi\-nimum values across
the halo)  and $V_s$ (maximum velocity shift measured across the halo)
  in Table 2 
refer to HeII, which is a more reliable kinematic tracer than Ly$\alpha$. 
 Ly$\alpha$ was used 
only for  2142-242, since this was the 
only line  detected from the quiescent
halo (\S3.1.7).

The Ly$\alpha$ emission of the 
quiescent halos is clearly detected  
in the outer regions of all objects 
except  0731+193 and 2105+236 (see top panels Fig.~1-9).
 The surface brightness  values presented in Table 2
 were measured using these
regions. 	It is important to note that the surface brightness 
varies spatially and it is often  higher in the inner apertures.
For  0731+193  and 2105+236 
 the surface brightness was estimated  using the 
flux of the narrow component isolated
with the kinematic analysis and the area of the apertures where this
component was detected (see\S3.1.5). 

	The  Ly$\alpha$ luminosities
  were calculated integrating the line flux
 from the quiescent
gas across all the apertures where this was isolated. 
 For 
0943-242 it was not possible
to isolate a narrow component using Ly$\alpha$
on the high surface brightness regions (see \S3.1). In this case
we show the luminosity of the HeII narrow line.

\subsection{Notes on individual objects}

\subsubsection{1809+407}

Narrow band H$\alpha$+[NII] imaging 
obtained by Egami et al. \shortcite{egami03} using
 NICMOS on HST
shows that the emission line nebulosity is misaligned from the radio axis. 
Our 2D Ly$\alpha$ Keck spectrum (Fig.~1) shows that in
addition to the high surface brightness regions, 
a LSBH with apparently quieter   kinematics 
(narrower Ly$\alpha$) is detected
extending $\sim$9 arcsec ($\sim$80 kpc)
  and beyond 
the radio structures.

Spectra were extracted from 5 apertures (Fig.~1).
The fits to HeII  reveal the presence of  a narrow component with
FWHM in the range $\sim$500-700 km s$^{-1}$ and maximum 
velocity shift across the nebula of $\sim$400 km s$^{-1}$. These results are
consistent within the errors with those obtained 
 using Ly$\alpha$.
Our results are in very good agreement with Egami et al. \shortcite{egami03}. These authors
found that the integrated near infrared (optical rest frame) emission lines can be
characteri\-zed by two components of FWHM$\sim$560 km s$^{-1}$ and 
1670 km s$^{-1}$ respectively  and shifted by $\sim$-470 km s$^{-1}$.

\subsubsection{0211-122}

 The 2D line spectra (Fig.~2, top) 
  show evidence  for a low surface brightness 
 halo which extends in Ly$\alpha$
for $\sim$13 arcsec  or $\sim$114 kpc.
The apparent kinematics of the LSBH seems rather 
uniform across its entire extension and quieter than in the high 
surface brightness regions.

Spectra were extracted  from  six apertures (Fig.~2). 
Both CIV  and HeII (except aperture 3, see below) are very well fitted with
single Gaussian profiles.  This is not always the case for Ly$\alpha$,
which is heavily absorbed in some positions \cite{ojik94} and therefore
not a reliable  kinematic tracer in this object.
The fits to the line profiles 
confirm that  the kinematics is quieter   in the outer parts 
of the object. Both HeII and CIV are unresolved 
(FWHM$\leq$400 km s$^{-1}$)  and  the
 velocity shift  ($\leq$ 100 km s$^{-1}$)  between these
two lines is small.

  Ly$\alpha$ is broad (FWHM$\sim$1000-1200 km s$^{-1}$) in the inner apertures
(ap. 3 and 5).  
 \footnote{Van Ojik measured FWHM(Ly$\alpha$) $\sim$ 1700 km s$^{-1}$ 
in the central 2 arcsec. The authors worked at 24 \AA ~  spectral resolution.}. 
Similar highly perturbed kine\-matics is not apparent in HeII, which shows
FWHM$\sim$700 km s$^{-1}$. 
The CIV ($\lambda\lambda$1548,1551)  profile 
is also consistent with two components of FWHM$\sim$700 km s$^{-1}$
and this is also the
case for the NV  ($\lambda\lambda$  1239,1243) doublet
(ap. 4) when broad underlying wings are taken 
into account.   We therefore conclude that the gas in the inner apertures
emits lines with FWHM$\sim$700 km s$^{-1}$, except for Ly$\alpha$. The maximum
velocity shift for the HeII line across the entire object is $\sim$200 km s$^{-1}$.

	It is interesting to note that NV is very strong relative
to HeII and CIV at a large distance ($\sim$45 kpc) 
from the nuclear region (see spectrum
ap. 6, Fig.~2). This will have implications on the chemical abundances
(see \S4).
 
\subsubsection{1931+480}

The Ly$\alpha$ image by Chambers et al. (1996) shows a centrally peaked 
nebula which
extends for $\sim$10 arcsec aligned with the radio axis.
The Keck spectrum (Fig.~3)
shows that the Ly$\alpha$ emission extends for $\sim$13.5 arcsec
($\sim$118 kpc). The 
low surface brightness emission shows apparently narrower Ly$\alpha$ 
 in the outer regions of the object.
Spectra were extracted from five apertures.

The fits to HeII reveal a 
narrow component in all apertures (2 to 5) 
with  FWHM in the range $\sim$450-720 km s$^{-1}$ and maximum
$V_s \sim$450 km s$^{-1}$  across the object.
The K+H image of these object shows four distinct clumps \cite{carson01}.  
 Near infrared KeckII
spectroscopy of the indi\-vidual clumps 
reveals that the FWHM of [OIII]$\lambda$5007
is   $\sim$525, 390, 840 and 570 km s$^{-1}$ for the individual
clumps.
These results are in 
good agreement with ours.

\subsubsection{1410-001}

 There is  a strong misalignment on this object between  the radio axis and the optical
structures as seen in the HST images ($\sim$45 deg,  Pentericci et al. 2001).
The slit was located along the radio axis, therefore 
the regions
 within the slit lie
 far beyond the HST structures (except maybe in the inner 2-3 arcsec). 

 The 2D Ly$\alpha$ spectrum  shows
low surface brightness emission in the outer parts
of the object with apparently quieter   kinematics (Fig.~4).

	Spectra were extracted from seven apertures.
 Ly$\alpha$ (ap. 7), HeII (ap. 5 and 6) and CIV
(ap. 6) are very na\-rrow (unresolved) in the outer parts of the object 
with FWHM(HeII)$<$472 km s$^{-1}$. HeII is broader ($\sim$700-800 km s$^{-1}$)
in the inner apertures. The maximum velocity shift across the object is
$\sim$500 km s$^{-1}$.
A narrower component 
is   suggested by the very 
narrow peak found on top of the CIV and  HeII  profiles at some
spatial positions.

\subsubsection{0731+438}

	Narrow band H$\alpha$+[NII] images obtained with the Subaru telescope
(Motohara et al. 2000) show diffuse line emission which extends out 
to $\sim$3.3 arcsec from the center. The
morphology is aligned with the axis defined by the radio hot spots. The authors interpret the observed
morphology as evidence for the ionization cones in this radio galaxy.
The Keck spectrum (Fig.~5)
 shows that  Ly$\alpha$ extends for $\sim$11 arcsec  ($\sim$96 kpc).

Three apertures were selected for the kinematic study (Fig.~5).
A narrow component  
 with FWHM$\sim$600 km s$^{-1}$ and $V_s\leq$100 km s$^{-1}$
is detected  in apertures 2 (Ly$\alpha$ and HeII)
and 3 (Ly$\alpha$, CIV and HeII).
The emission from aperture 1 corresponds to a separate component clearly
seen in the 2D spectra which emits broad lines (FWHM$\sim$1000 km s$^{-1}$). 
We have not considered this component as part
of the quiescent LSBHs.
 The results obtained with the three emission lines are in good agreement.

\subsubsection{2105+236}

Ly$\alpha$ narrow band imaging shows two opposing ionization cones 
 with $\sim$90$\degr$
opening angles  (Knopp \& Chambers 1997).
This is one of the rare cases among high redshift radio galaxies 
with clear ionization cones.
The 2D Keck spectrum (Fig.~6, top)
 shows emission from these two spatial components in all
emission lines.
 Contrary to the rest of the objects in this sample,
 the 2D spectrum does not
show clear evidence for a quiescent LSBH.

Spectra were extracted from six apertures (Fig~6).
The fits both to Ly$\alpha$ and HeII  are rather complex in this object
due to the presence of an underlying very broad component (probably scattered
light, Vernet et al. 2001) detected across the whole extension of the SW component,
which makes the setting of the continuum level uncertain. We have attempted
to fit the main peak of the emission line as accurately as possible, rather than
the broad under\-lying wings, which are difficult to reproduce. The errorbars
in Fig.~6 (bottom right panels) account  for the different possi\-bilities considered as the continuum
level for the fits. This uncertainty is large for the high velocity component,
but it does not affect  seriously the narrow component, at least in HeII.

HeII is  narrow in all apertures (2 to 6) with  
FWHM$\sim$500-650 km s$^{-1}$ and the maximum velocity shift across
the nebula is $\sim$570 km s$^{-1}$. CIV is also narrow in ap. 5 with 
FWHM$\sim$600 km s$^{-1}$ and similar velocity shift. 

	The fit to  Ly$\alpha$  reveals  the presence of
a broad component (FWHM$\sim$1800 km s$^{-1}$) and  a narrower component 
with FWHM$\sim$500-1100 km s$^{-1}$. Except in ap. 5, both 
 the FWHM and velocity shift
 of this component  are consistent within the errors with the 
 HeII measurements.

\subsubsection{2104-242}

The 2D VLT spectra (Fig.~7) show emission from two
spatial components that where described for the first time by McCarthy et al. 
\shortcite{mac90b}. 
These two clumps are between and aligned with the radio lobes. 		
The Ly$\alpha$ emission extends for $\sim$15 arcsec ($\sim$130 kpc), 
well beyond the structures
seen in the Ly$\alpha$ image (Pentericci et al. 1999). 
 Very narrow Ly$\alpha$ is detected towards the north-east.

	Spectra were extracted from 6 apertures.
The kinematic structure (see also Overzier et al. 2001)
of this object is very complex, but recall that the spectrum has higher spectral
resolution than for the rest of the sample. 
Two narrow (FWHM(HeII)$\leq$200 and  $\sim$400 km s$^{-1}$
respectively) components are found in the southern clump, 
which is  associated with the narrow filament
more than 2 arcsec long detected in  HST images (Pentericci et al. 1999).

An interesting feature is the
 very narrow Ly$\alpha$ emission (FWHM$\leq$265 km s$^{-1}$) 
 detected in the outer regions of
the object (ap. 1 and 2). 
The fit to the Ly$\alpha$ profile in ap. 6 reveals a similar 
component (similar FWHM and velocity shift) 
also in
the outer regions of the object to the south. 
Absorption of Ly$\alpha$ photons could be responsible for the narrowness of the line,
but
 the similarity of the kinematic structure of all the lines
 shown by the 2D spectra and the large strength of Ly$\alpha$ relative
to other lines  indicates that absorption is not important in this object.

It is uncertain whether the outer very narrow line LSBH 
is also detected across the 
high surface brightness regions. 
Our spectra
do not reveal a component of similar FWHM and velocity shift but  it might be 
concealed in the
two components of the southern clump. Higher resolution (2.8 \AA) spectroscopy reveals three components (rather than 2) with
velocity shifts of 140, 670 and
1270 km s$^{-1}$ respectively (Pentericci et al. 2001;
 the authors do not provide
FWHM va\-lues). One of these three components might correspond to the 
outer LSBH.
In Table 2, we have assumed that the quiescent LSBH is the
very narrow line  emitting
gas  in the outer parts of the object.

\subsubsection{1558-003}

Low surface brightness Ly$\alpha$
emission is detected beyond the radio structures extending for
$\sim$14 arcsec ($\sim$120 kpc) with apparently narrower lines
(Fig.~8, top). 
Spectra were extracted from 6 apertures. 
 The fits reveal a narrow component with 
FWHM in the range $\leq$550-630 
km s$^{-1}$ in Ly$\alpha$ (ap. 1), HeII (ap.2) and CIV (ap. 5).
The maximum velocity shift for HeII across the nebula is $V_s\sim$550 km s$^{-1}$.

 HeII is broader in the central regions  (FWHM$\sim$750-850 km s$^{-1}$)
but  the presence of a narrow component is sugges\-ted by the very 
narrow peak found on top of the CIV and HeII  profiles (ap. 4).  
Although the errors are large ($\sim$40\% for the FWHM$_{obs}$ of the narrow
component) the spectral profile of HeII
in ap. 3 and 4 is  well 
 reproduced by  a narrow  unresolved component with 
FWHM$\leq$410 km s$^{-1}$ and $V_s\sim$+230 and +340 km s$^{-1}$
respectively,  and a broad component 
of FWHM $\sim$1000 km s$^{-1}$. (Because of the large errors,
we have ignored the results of these fits in Table 2).

It is interesting to note that broad Ly$\alpha$
 (FWHM$\sim$1200 km s$^{-1}$)
 is found beyond the radio structures. This was also reported by Villar-Mart\'\i n,
 Binette \& Fosbury (1999).

\subsubsection{0943-242}

Clear evidence 
for a low surface brightness 
very extended halo with apparently quiet  kinematics
is found in the 2D spectrum of all (Ly$\alpha$, NV, CIV and HeII) lines
 (Fig.~9, top). This gas 
extends for $\sim$8 arcsec (67 kpc) from the continuum
centroid, well beyond the radio structures.

Spectra were extracted from 5 apertures. 
The double peaked profile found in some of them  
is due to absorption (e.g. R\"ottgering et al. 1995). Since the effect is so dramatic on
the profile of the line, we have focussed the kinematic analysis on the HeII line
instead. 

	A narrow component of FWHM$\sim$400-600 km s$^{-1}$ is found in 
HeII in all apertures with maximum velocity shift  $\sim$450 km s$^{-1}$
across the nebula. The
results for CIV are in very good agreement. An interesting feature 
 is the strength of the NV line in the quiescent halo relative to
other lines (Fig.~10).

\subsection{Seeing disk and slit width: Uncertainties}

So far we have assumed that the objects filled the slit.
In some cases, the seeing disk was smaller than the slit width
and our assumption introduces additional uncertainties. 
On one hand, if the objects are clumpy,
 the image of a spatially unresolved  clump may be smaller
than the slit width and the instrumental profile at that particular
position is narrower than the profile measured with the sky or arc 
lines. This would lead to an underestimation of the intrinsic FWHM
of the emission lines. 
On the other hand, it is possible that some clumps
may be displaced in the dispersion direction relative to the center
of the slit. The net effect will be  errors in the velocity
fields since these  may partially reflect the distribution of the
emitting gas within the slit.

	We do not expect these uncertainties to affect the results 
on  0731+438, 0828+193, 1809+407, 1931+480 and 2104-242. 
For 0731+438, 0828+193 and 2104-242 we are confident that 
the  seeing disk filled the slit. 
For 1809+407 and 1931+480, it is reassuring that  
our results  are 
in  good agreement with Egami et al. \shortcite{egami03} and 
Carson et al. \shortcite{carson01}
respectively. 

For the rest of the sample  at least
part ot the data where obtained  with seeing size smaller than
the slit width. 

We have estimated the maximum uncertainties on both the FWHM
and $V_s$ values assuming the worse case scenario, 
i.e., the best seeing  measured during the observations
for each object and a clumpy morphology.
The uncertainties estimated in this way are therefore  
upper limits. The ins\-trumental profile
for a spatially unresolved source would be 
$\frac{seeing~size~(arcsec)}{slit~width~(arcsec)}\times$IP, where IP 
is measured using the sky  lines (Table 1). The range of FWHM
we obtain across the halos are: $\leq$550-790 km s$^{-1}$ for 0211-122;
540-700 km s$^{-1}$  for 0943-242; $\leq$710--960  km s$^{-1}$ for 1410-001;
680-940 km s$^{-1}$ for 1558-003; and 700-810 for 2105+236.
It could also be that some of the emission lines which we
have found to be unresolved turn out to be resolved.

	We estimate that
the maximum velocity shift (for HeII) which could be artificially 
introduced 
due to the spatial displacement
 between  unresolved clumps in the direction perpendicular to the slit
 would be $\sim$310-365 km s$^{-1}$ (depen\-ding on the object)
for  0211-122, 0943-242 and 1558-003 and 
$\sim$600 km s$^{-1}$ for
 1410-001 and 2105+236. 

	Since all the estimations above have been done for the worse case
scenario, we are confident that 
our main conclusion relative to the existence of quiescent
gas
in all objects are not  
affected. Our assumption that the objects 
filled the slit 
 could have some non-negligible
impact on some of the calculations in \S4.4. This will be mentioned
when appro\-priate.

\section{Discussion}

	In this section we will investigate  the origin of the 
velocity patterns, the ionizing mechanism, 
and the nature  of the LSBHs as well as the  cosmological 
implications of our results.

\subsection{Nature of the  kinematic components}

	There is evidence that the kinematic and ionization properties of high
redshift radio galaxies can be determined in va\-rying degrees by shocks 
driven during interactions
between the radio structures and the ambient gas as well as photoionization by
the central quasar (e.g. Best, R\"ottgering \& Longair 2000).  
This is also the case of some intermediate redshift radio galaxies.
As an example, Villar-Mart\'\i n et al. \shortcite{vill99}
explained the kinematics and optical emission line spectrum of the radio galaxy
PKS2250-41 ($z=$0.3) in terms of two kinematic components one of which is emission
from kinematically perturbed shocked gas and the other is non perturbed gas 
photoionized by the continuum from the active nucleus or the continuum generated
by the shocked hot gas.

We have found that all objects in our sample are asso\-ciated with
giant gaseous reservoirs with quiescent kinema\-tics that sometimes extend
beyond the radio structures.
In addition, we have found highly perturbed gas in 8 out of 10 
objects in the sample, usually located inside the radio structures.
We propose that these are the emission from
ambient non shocked gas  and the  emission from shocked gas respectively.

	There are other examples in the literature
 that support our interpretation.
Spectroscopic observations of 4C41.17 ($z=$3.8, van Breugel et al. 2001)
and  1243+036  \cite{ojik96}  
show that the gas kinematics is highly
perturbed  inside the radio structures and along the radio axis. 
Beyond the radio hot spot 
the velocity shifts and FWHM of the lines decrease abruptly.
See also Maxfield et al. \shortcite{max02}.

	The discovery of a quiescent reservoir of gas in all the objects
in our sample  suggests that it might be a common
ingredient of high redshift radio galaxies. 

\subsection{Properties  of the quiescent LSBHs}

\subsubsection{Sizes}

Baum et al. (1988) and Baum \& Heckman (1989) studied the 
emission line properties of a sample of 43 low redshift ($z<$0.6)
powerful radio galaxies. In $\sim$85\% of these sources the line
emission was  spatially resolved. They found that the median total extent
of the emission line nebulae is 10 kpc, with only one object (3C227) with size (96 kpc)
larger than the minimum total extension of the  quiescent halos in our
sample ($\sim$60 kpc; the maximum total extension is $\sim$140 kpc). 
The LSBHs are  giant structures.

 There are  a few examples in the literature of HzRG
with Ly$\alpha$ emission on
the  100 kpc scale (McCarthy et al. 1990a, Francis et al. 2001, 
Pentericci et al. 1998,
 Reuland et al. 2003b). Our work shows that 
all the objects in our sample are associated
 with giant 
 halos which
extend far beyond the
continuum structures revealed by the HST images and the high surface
brightness emission line regions.

\subsubsection{Luminosities}

The  H$\alpha$ luminosities of the quiescent halos  within the slit
expected from the measured Ly$\alpha$ luminosities
 (assuming case B recombination Ly$\alpha$/H$\alpha$=8.8)
are in the range $\ga$0.1-2.8$\times$10$^{43}$ erg s$^{-1}$
(these are lower limits because Ly$\alpha$ absorption has been neglected).
Baum \& Heckman (1989) found that the median emission line
luminosity in H$\alpha$+[NII] in their low redshift sample
 is 3$\times$10$^{41}$ erg s$^{-1}$.
Since the halos are likely to occupy a much larger spatial
region outside the slit, it is clear that the total luminosities are 
 probably   much
higher
than line luminosities in low redshift radio galaxies.

We have estimated the total Ly$\alpha$ luminosity 
expected from the halos, assuming that the gas is
 photoionized by the quasar continuum (see \S4.3.1)
 We emphasize the
 approxi\-mate nature of these
calculations due to the strong assumptions relative
to the geometry of the ionized gas, the neglection of Ly$\alpha$
absorption  and   the constant value of the Ly$\alpha$ luminosity
per unit volume across the halo.
A biconical geo\-metry  with
cone opening angle  90$\degr$ is naturally expected in this scenario (Barthel 1989).
We have assumed that the two cones have convex base and heights given by the extension
of the LSBHs at both sides of the continuum centroid.
The Ly$\alpha$ luminosities (Table 2) and the volumes 
of the LSBHs within  the slit  and within the  ionization
cones were used
to estimate an average luminosity  per
unit volume and  the total luminosity expected from the two cones. 
We obtain
$L^{tot}_{Ly\alpha}$ in the range 7.7$\times$10$^{43}$ - 
1.2$\times$10$^{45}$ erg s$^{-1}$.
For comparison, Reuland et al. (2003b)
measured  total
 Ly$\alpha$ luminosities of the giant nebulae in 3 radio galaxies
at $z\sim$3.4-3.8 in the range 5.9$\times$10$^{44}$-1.4$\times$10$^{45}$ erg s$^{-1}$.
These values include the emission from the high surface brightness perturbed regions.
This suggests that the quiescent gas emits a large fraction of the total
line luminosity  in HzRG.

The implied H$\alpha$ luminosities are  
$\ga$8.7$\times$10$^{42}$-1.4$\times$10$^{44}$ erg s$^{-1}$,
i.e., about  $\ga$100-1000 times higher than
those of low redshift powerful radio galaxies (Baum et al. 1989). 

\subsubsection{Kinematics}

The maximum velocity shift measured across the total extension of the
quiescent halos in our sample is $\leq$600 km s$^{-1}$ and 
 the
 maximum FWHM, except for 1410-001 and 1558-003,
is $\sim$700 km s$^{-1}$ (Table 2). In those two radio galaxies, the
maximum FWHM is $\sim$800-850
km s$^{-1}$. It is important to note that 
spectral decomposition of HeII was not attempted in these
two objects, but in both cases the presence of a narrower component
 is expected (see \S3.1).

Baum et al. (1990) studied the kinematic properties of the extended gas
in a sample of 19 low redshift powerful radio galaxies. 
Emission line FWHM and maximum velocity shifts across the 
extended
 nebulae 
are in general $\la$500 km s$^{-1}$.
At low redshift, although rare, there are some 
objects where  velocity shifts as high as
$\sim$800 km s$^{-1}$ have been measured  and which present
 regular motions 
consistent with rotation (e.g. Tadhunter,
Fosbury \& Quinn 1989).

Although the lines in the HzRG halos are somewhat broader
the FWHM values are not extreme and do not imply kinematic perturbation.
FWHM larger than 600 km s$^{-1}$ are very rare in the extended
nebulae (Baum et al. 1990) of low 
redshift radio galaxies, but some examples with no
signs of kinematic perturbation exist: for instance, 
PKS 1345+125, which was classified by Baum, Heckman \& van Breugel (1992)
as a `rotator'. 

The broader lines in our sample might also be  a consequence
of the larger linear size covered by the apertures used in our kinematic 
analysis ($>$10 kpc, compared with $<$1 kpc in Baum, Heckman \& 
van Breugel (1990)
study).  Velocity shifts $>$200 km s$^{-1}$ across less
than 10 kpc are common in low redshift
radio galaxy nebulae. Such shifts will broaden the line profiles when 
integrating over larger (linear scale) apertures.
As a reference for comparison, the typical line-of-sight {\it stellar}
velocity dispersion of a giant elliptical galaxy is $\sigma_{los}\sim$300
km s$^{-1}$, corresponding to a characteristic line width (FWHM) of
$\sim$700 km s$^{-1}$ (e.g. Heckman et al. 1991a). 

	Therefore, the velocity shifts  across the 
quiescent halos are consistent with measurements in low redshift radio
galaxies. The lines are   often somewhat broader,
 but the FWHM values are not extreme and consistent with quiet
(non perturbed) kinematics.

\subsubsection{Densities}

Electron densities $n_e$ for the quiescent halos were  estimated 
using the relation  $L_{Ly\alpha}$ = 4 $\times$ 10$^{-24}~ n_e^2~f~ V$ erg s$^{-1}$, 
where $f$ is the filling factor, and $V$ the
volume of ionized gas inside the slit (McCarthy et al. 1990a) The same geometry as in
\S4.2.2 was assumed: $V$ is the slice of the convex
cones included in the slit.  Large uncertainties affect the calculations 
 due to the assumptions on the geometry, the neglection
of Ly$\alpha$ absorption effects,  the uncertainty on the 
$f$ values and the assumption
that the density is constant across the halo.
$f$ values have been estimated to be $\sim$10$^{-5}$ $\sim$10$^{-4}$ for 
typical emission line regions of HzRG
\cite{mac93} \footnote{$f$ values in low redshift radio galaxies
(where the density can be constrained using the [SII] doublet)
are in the range 10$^{-6}$-10$^{-4}$  
(Heckman et al. 1984, van Breugel et al. 1985)}.
However, $f\sim$ values in the range 10$^{-7}$-10$^{-8}$ 
were obtained in  studies 
of Ly$\alpha$ nebulosities associated with high  redshift quasars 
(Heckman et al. 1991b). 
The advantage in this case is that 
one knows the properties of
the ionizing source with some confidence.

 We obtain densities in the range $\sim$17-150 cm$^{-3}$ for
$f=$10$^{-5}$ (a factor of 10 higher if $f$=10$^{-7}$). 
 In spite
of all the sources of uncertainty, the calculated
densities are similar to estimations by other authors 
using other techniques (e.g. Heckman et al. 1989, Heckman et al. 1991b) 
for gaseous ne\-bulae
associated with high redshift radio galaxies and quasars. As an example, 
McCarthy et al. \shortcite{mac90a}
calculated an {\it upper} limit for the density of the ionized
 gas in 3C294 of $\sim$150 cm$^{-3}$ based
on estimations of the pressure in the diffuse radio plasma.

\subsubsection{Masses}

The masses are calculated as $M = n_e ~ V ~ f ~m_p$, 
where $m_p$ is the proton
mass and $V$ is the volume within the ionization cones.
For the densities derived above and assuming the same biconical geometry
and $f=$10$^{-5}$, 
the mass of ionized gas in the quiescent
halos  is in the range 10$^{9-10}$ M$_{\odot}$ (a factor of 10 lower
for $f$=10$^{-7}$).  For comparison, McCarthy (1993) quoted
typical masses of ionized gas
in HzRG of $\sim$10$^9$ M$_{\odot}$.


\subsection{LSBH energetics}

 We 
investigate in this section the excitation mechanism of the giant LSBHs:
photoionization by the quasar con\-tinuum, photoionization by the continuum 
generated by the hot shocked gas, photoionization by stars, 
cooling radiation from a cooling flow nebula.

\subsubsection{Quasar}

We estimate in this section the total quasar ionizing luminosities
 $Q^{tot}_{ion}$ implied by the Ly$\alpha$ luminosities 
of the quiescent halos.
The total  Ly$\alpha$ luminosities expected
from the ionization cones were 
calculated in \S4.2.2. 
We have estimated $Q^{tot}_{ion}$ assuming that in case B
recombination 68\% of hydrogen ionizations yield a Ly$\alpha$ 
photon (Osterbrock  1989). We have also assumed a 
covering factor $C=$0.15 
of the quies\-cent halos as seen by the nucleus (Heckman et al. 1991a).
We obtain  $Q^{tot}_{ion}$ in the range
3.1$\times$10$^{55}$ -7.0$\times$10$^{56}$  photons s$^{-1}$. 

If the ionizing continuum follows a power law of index
 $\alpha$=-1.0 \cite{vill97},  the total ionizing luminosities  
are in the range  4.4$\times$10$^{45}$-10$^{47}$ 
erg s$^{-1}$.  
In spite of the
large uncertainties involved in the calculations, it is
interesting to note that the values we have obtained are
in the same range as radio loud quasars (McLure \& Dunlop 2001) 
at comparable redshift \footnote{High redshift quasars ($z=$2-3.4) 
with luminosities
as high as 10$^{48}$ erg s$^{-1}$ have been reported 
(Kaspi et al. 2003)}.

	We conclude that the halo luminosities and measured sizes 
can be naturally explained by quasar photoionization.
This is further supported by the emission line spectra (strong CIV, HeII and in some
cases NV emission) typical of active galaxies, and inconsistent with
 star forming objects.

\subsubsection{Shocks}

 We have calculated the UV ionizing luminosity $L_{ion}^{shock}$ ge\-nerated by  shocks of different
velocities (range 500-900 km s$^{-1}$, Bicknell et al. \shortcite{bick00})
which can explain the measured Ly$\alpha$ luminosities $L_{Ly\alpha}$
 of the LSBHs \footnote{The quiet  kinematics of the halos 
suggests that this gas 
has not been shocked, therefore, we reject collisional ionization.}. 	
Combining equations (4.1) and (4.4) in Dopita \& Sutherland (1996)
and assuming case B recombination value for Ly$\alpha$/H$\beta$=26.5:

$$L_{ion}^{shock} = 4.2 ~\times~ (\frac{V_s}{100~km~s^{-1}})^{0.84}~\times~ L_{Ly\alpha} ~~ 
erg ~s^{-1} $$

The total luminosity radiated by the shock is $\sim$2$\times L_{ion}^{shock}$ (Dopita \& Sutherland 1996)
and a large fraction should be emitted within the optical filters. 
Using the Ly$\alpha$  luminosities
measured within the slit for the objects in our sample  the resulting $L_{ion}^{shock}$
values are in the range $\sim$several $\times$ 10$^{44-45}$ erg s$^{-1}$ for 500 
km s$^{-1}$ shock velocity (higher values are obtained for higher velocities). 

 There is, however, no evidence for such source of continuum in the extended regions
of HzRG. The high velocity gas   gives us an idea of  the spatial extent of the 
shocked gas. According to this, the continuum   should
have surface brightness in the range $\sim$several $\times$ 10$^{-14}$-10$^{-15}$ 
erg s$^{-1}$ cm$^{-2}$ arcsec$^{-2}$. This is a gross lower limit, since we have
only considered the gas within the slit. Typical optical 
surface brightness  of HzRG measured
from WFPC2/HST images are however $\la$few $\times$ 10$^{-16}$ 
erg s$^{-1}$ cm$^{-2}$ arcsec$^{-2}$.

	Another argument against shock processes is the extreme mass flow rates
 (\.M) through the shock required
to explain the Ly$\alpha$ luminosity (Villar-Mart\'\i n et al. 1999). The required values
are in the range 10$^{3-4}$ M$_{\odot}$ yr$^{-1}$ for $V_s$=500 km s$^{-1}$ 
(a factor 2.1 smaller for 
$V_s$=900 km s$^{-1}$). If the age of the radio source is $\sim$10$^7$ yr 
and it is entraining material
during its pass through the gaseous environment, this would imply
that the total amount of material `consumed' by the shock  is  in the range
10$^{10-11}$ 
M$_{\odot}$. These masses are larger than  
the total mass of ionized gas in high redshift radio galaxies 
(e.g. McCarthy et al. 1990a, van Ojik et al. 1997, Villar-Mart\'\i n et al. 2002).
The amount of `consumed' gas is likely to be  higher, since we are only considering the
one required to produce the luminosity measured inside the slit.

It is therefore unlikely that the  excitation of the quiescent halos
is due to  shock related
processes.

\subsubsection{Cooling radiation}

  Steidel  et al. (2000)
 proposed that the  Ly$\alpha$
giant ($\ga$100 kpc) nebulae  found around Ly break galaxies at $z\sim$3  
can be powered by radiative
cooling of gas.  Could this be the origin of the emission from the
quiescent 
LSBHs in our sample?

	Models of high redshift cooling flows 
for galactic systems
in the process of formation \cite{fardal01} predict
 Ly$\alpha$ luminosities due to cooling radiation as
 high as few $\times$ 10$^{44}$ erg s$^{-1}$.
According to these models the range of measured 
Ly$\alpha$ luminosities from the quiescent halos  (Table 2)
imply star forming rates (SFR) in the range $\sim$50-700 M$_{\odot}$.
These are gross lower limits, since
we have used only the luminosity within the slit and neglected
Ly$\alpha$ absorption effects.

	 Using the luminosity (within the slit) at 1500 \AA\, 
Vernet et al. \shortcite{ver01} estimated a range in SFR of 2-60 M$_{\odot}$ yr$^{-1}$
for  the objects  in our sample,  not taking into account
dust reddening. Although the estimations are affected by large
uncertainties, it is interesting to compare with the model predictions
by Fardal et al (2001). If 
the ratio of un-obscured to obscured SFR
 is similar to that of Lyman
break galaxies at $z<$3 (i.e. $\sim$25, Sawicki \& Yee 1998),
 the  
SFR in our objects could be in 
the range 50-1500 M$_{\odot}$ yr$^{-1}$. If this is the case
and according to Fardal et al. models, it is possible that 
in some cases the measured Ly$\alpha$ luminosities
are consistent with cooling radiation.

	A problem for these models is that
the line emission  is expected to be concentrated in a very
small region of radius $\la$10 kpc or less for the most luminous halos
 ($\geq$few $\times$ 10$^{43}$ erg s$^{-1}$). To explain the 
large sizes of the Ly$\alpha$ nebulae 
around Ly break galaxies discovered by Steidel et al. (2000), the authors propose that the Ly$\alpha$ photons
from  gravitational cooling are scattered by neutral hydrogen in the
outer neutral regions.
However, we reject this interpretation since we have detected
HeII and CIV in the outer parts of the quiescent halos and therefore,
this gas is ionized.

	Similar cooling flow models  by
 Haiman, Spaans \& Quataert \shortcite{hai00} predict
 Ly$\alpha$ luminosities in the range 10$^{43-44}$ erg s$^{-1}$ due to cooling 
radiation within measured angular sizes in the range
11 to 15 arcsec for $z=$2 to 3, consistent with the $L_{Ly\alpha}$ values in Table 2. 
 One of the important differences with
Fardal et al. \shortcite{fardal01} models is the assumption that
Ly$\alpha$ is produced out to the virial radius, while Fardal et al. (2001)
 predict a much  more concentrated Ly$\alpha$ emission.
A problem for Haiman, Spaans \& Quataert \shortcite{hai00} models is that 
the expected surface brightness
for the halos is
$\sim$few $\times$ 10$^{-19}$ erg s$^{-1}$ cm$^{-2}$ arcsec$^{-2}$, 
a factor of at least 100 lower than we observe. 

	In summary, it is unlikely that the line emission from
the giant quiescent halos is due to cooling radiation.
Some high redshift cooling flow models
can explain the measured 
Ly$\alpha$  luminosities  within the slit, 
however 
the models are
inconsistent with the measured surface brightness or the angular
sizes of the halos. In addition, a cooling flow nebula is likely to occupy
a much larger volume than that within the slit and in such case,
the high  expected Ly$\alpha$ luminosities would  be also difficult to
reconcile with cooling flow models.

\subsubsection{Stars}

Continuum has been detected from some of the LSBHs, which 
 could be due to stars. We cannot
confirm this, however. Scattered light or nebular continuum might be  dominant
(Vernet et al. 2001). 
A strong argument against stellar photoionization 
is the detection of high ionization lines such as CIV, NV and HeII,
 which require
a much harder ioni\-zing continuum.
The LSBH emission line spectra  are  typical  of high redshift
active galaxies and very different from  high redshift star forming objects.

\vspace{0.5cm}

Therefore, we conclude that the ionization of the quiescent giant halos
 along the radio axis is   dominated by the
quasar continuum.
We do not reject the possibility that other mechanisms  excite
 the gas  in other regions across 
the LSBHs.
The giant  Ly$\alpha$ nebulae revealed by narrow band images
obtained by Reuland et al. (2003b) show  
 Ly$\alpha$ emission in regions which are  also
 outside any putative ionization cones.  Unless Ly$\alpha$ is 
scattered (Villar-Mart\'\i n, Binette \& Fosbury, 1996) another mechanism 
must be
 responsible for the excitation of this gas.

\subsection{Origin of the halos and dynamical masses of HzRG}

\subsubsection{Are the halos settled in a stable configuration ?}

\centerline{\it Rotating disks }

	VO96 and VM02 proposed that the giant quiescent halos could
be settled in rotating disks. Both observations and theory \cite{barnes02} indicate
that
much of the interstellar gas in merging galaxies may settle into 
gaseous disks which may extend to several times the remnant half-light radii.
 Rotation of the extended emission
line gas in some powerful radio galaxies at low redshifts suggests the presence
of young disks of gas acquired in a recent interaction or merger with a gas
rich galaxy \cite{baum92}. 
	Large scale (up to 100 kpc) HI disk like  structures 
have been found in several low redshift radio galaxies
\cite{mor02} and elliptical galaxies \cite{oo02}.

	We have investigated whether there are signs of
rotation in the LSBHs of our sample of HzRG.

\begin{itemize}

\item	3 objects  (0211-122,  1558-003 and 1931+480)
show velocity curves inconsistent with rotation.

\item	 4 objects  (0943-242, 1410-001, 2104-242, 2105+236) show velocity
fields consistent with  rotation curves with the slit
 located along or close to 
the line of nodes. For 1410-001, the velocity shifts at the positive
side of the spatial zero might map half
of the rotation curve (Fig. 4, bottom)

\item   1 object (1809+407) shows a velocity field
consistent with a rotation curve
with the typical V shape expected when the slit is located perpendicular
to the line of nodes and with some impact parameter relative to the
rotation center. This is also the case of 0828+193 (VM02)

\item for 0731+438 it is not possible to say, 
since the quiescent gas
is detected only in two apertures. 

\end{itemize}

	Therefore, 6  (including 0828+193) out of 10 objects in the sample have velocity
fields consistent with  rotation. The rotation center (probably located at the position
of the active nucleus) is clearly shifted from the 
spatial continuum centroid in some objects. 
This is expected, since the optical continuum
is not direct AGN light, but nebular, stellar and/or scattered quasar light.

When the rotation curve is measured along the line of nodes,   the
dynamical mass can be calculated as  $M_{dyn} = \frac{R~V^2}{G~sin^2~i}$,
where $R$ is the radius of the disc, $V$ is half 
the amplitude of the rotation curve and $i$ is the inclination
angle of the disk with respect to the plane of the sky.  
We have calculated dynamical masses for
0943-242,  2104-242 and 2105+236.
$R$ is the maximum extension of the LSBH from the 
rotation center. This was determined for 0943-242 from
 the rotation curve 
(Fig~9, bottom).  We have assumed that this position is
marked by the radio core in   2104-242 and 2105+236.
This is a good assumption for 2105+236, since 
the radio core was aligned with the vertex of the ioni\-zation
cones (\S2.2), where the active nucleus is expected to be.
For 2104-242 the assumption might be wrong, since the
NIR continuum centroid (which was aligned with the radio core)
might be spatially shifted from the AGN. The results, however, are
not strongly affected when considering a reasonable range of possible
positions for the rotation center.

We obtain $M_{dyn}\times sin^2i$ 
$\sim$2.8, 0.3 and 
2.9$\times$10$^{12}$ M$_{\odot}$ for 0943-242, 2104-242 and
2105+236 respectively\footnote{If the gas across 0943-242 and 2105+236 did not fill the slit, 
 the velocity curves might partially reflect the gas distribution
within the slit rather than rotation (see \S3.2).}.
For comparison, vO96 estimated 
$M_{dyn}\times sin^2i\sim$10$^{12}$ M$_{\odot}$ for 1243+036.
	Typical masses of cD galaxies at low redshift 
are $\sim$10$^{13}$ M$_{\odot}$
or more. If the radio galaxies above 
have similar masses, this implies
 disk inclination angles $i\la$30$\degr$  relative
to the plane of the sky.

	Alternative possibilities that would produce shallow rotation curves are
1) locating the slit across the rotation center and 
at certain angle ($<$90 degr)  relative to the line of nodes 
2)  the objects  are less massive than  cD giant ellipticals at
low redshift. This result was also obtained by Tadhunter,
 Fosbury \& Quinn \shortcite{tadh89} from
the kinematic study of the extended gas in low redshift powerful radio
 galaxies.

\vspace{0.3cm}

\centerline{\it Companion satellites}

\vspace{0.2cm}

As VM02 proposed, the halos might consist of
a number of satellite companions. There is evidence that powerful radio galaxies 
at low redshift lie in rich 
environments, with numerous companions at small distances from the main galaxy
(within the inner 100 kpc). Some examples are 3C171 
\cite{heck84}, PKS2250-40 \cite{tadh94} and Coma A \cite{tadh00}. 
HzRG are known to lie in rich environments (Pentericci et al. 2002a, Venemans
et al. 2002) and numerous star forming compa\-nions have been found in the field
of at least one high redshift radio galaxy \cite{pente02a}.

	However, the continuous detection of the quiescent 
halos across tens of kpc and the AGN type spectrum
su\-ggests that they are gaseous reservoirs,
rather than companion  (proto)galaxies. 
Although we could think 
of a scenario where the central
(proto)galaxy  is surrounded by a high number of active galaxies 
\cite{pente02b}
 or that the central quasar ionizes the gas in the companion galaxies, 
a  more natural explanation is that the halos are  gaseous reservoirs.
The diffuse and chaotic (filaments, plumes, cones)
 morphologies of the giant Ly$\alpha$ nebulae revealed
by recent  narrow band  images of high
redshift radio galaxies \cite{reu03b} further support this interpretation. 
Companion satelli\-tes might be embedded in the diffuse halos
but the line emission is likely to be dominated by the  halos. 
Reuland et al. \shortcite{reu03b} find `gaps' in the giant 
Ly$\alpha$ nebulae which coincide
with companion satellites revealed by K band images.

\vspace{0.3cm}

\centerline{\it Gaseous spherical envelopes}

\vspace{0.2cm}

	Rather than settled in a disk, 
the quiescent  halos could be   gaseous envelopes supported 
against gravitation by the velocity dispersion within the halos (rather than rotation).
The dynamical mass enclosed within a radius $R$
 can be calculated as $M_{dyn} = \frac{5~R~V_R^2}{G}$,
where $V_R$ is in this case the radial velocity dispersion of the clouds 
\cite{carr99} 
within the halo (given by the $\sigma$ of the emission lines) and $R$ is $R_{max}$ (Table 2). 
The estimated masses  are shown in Table 3.

\begin{table}
\centering
\begin{tabular}{cccc}
\hline
 	  &  	$M_{dyn}^{min}$   & 	$M_{dyn}^{max}$  \\
 Name	&	within $R_{Max}$ & within $R_{Max}$ &  \\ \hline
1809+407    &  2.1  & 4.1\\
0211-122    &  $\leq$1.2 & 6.3    \\ 
1931+480    &  2.3     & 5.8\\ 
1410-001    &  $\leq$2.9 &  8.3  \\ 
0731+438    &  1.8     &  \\ 
2105+236    &  3.2   & 5.4  \\ 
2104-242    &  $\leq$1.0 &   \\ 
1558-003    &  4.4    & 10.0  \\ 
0828+193    &  $\leq$1.3 &    \\
0943-242   &  2.3  & 4.4 \\ \hline
1243+036    &  1.0  &   \\
\hline
\end{tabular}
\caption{Dynamical masses of high redshift radio galaxies assuming that
the  gas consists of a virialized system of clouds. The second and
third 
columns give the masses using the minimum and maximum FWHM of HeII
across the halos (Table 2).  Masses are given 
in units of 10$^{12}$ M$_{\odot}$. Upper limits are given for those
objects for which  FWHM$_{min}$   are upper limits.}
\end{table}

Therefore, M$_{dyn} \la$several $\times$ 10$^{12}$ M$_{\odot}$ for most
objects\footnote{The calculations for 0211-122,  0943-242, 1410-001,
1558-003 and 2105+236
could be affected by seeing effects (see \S3.2). 
Considering  the worse case scenario 
the masses would change   to 8.1, 7.1, 
12.1, 13.0, 8.5 $\times$10$^{12}$, M$_{\odot}$ respectively, using
the maximum FWHM measured across the quiescent halos (Table 2)}. 
If the halos are virialized systems of clouds,
these results suggest that most radio galaxies
 in this sample are less massive than
nearby cD giant elliptical galaxies.

\subsubsection{Inflows or Outflows}
  
	The possibility that the quiescent halos are cooling
flow nebu\-lae was discussed
in \S4.2. Although we concluded that the line emission is likely to
be powered by the quasar continuum, 
this does not exclude the possibility that the halos are cooling
flow nebulae (but with the quasar dominating the emission line processes).
This will be discussed in more detail in \S4.5.

	An alternative possibility 
is that the halo has  been deposited by galactic
superwinds.  Winds are needed to explain the chemical enrichment of the quiescent halos
 at tens of kpc from the nuclear region (the strength of the NV
line relative to CIV and HeII in 0211-122 and 0943-242 suggests at least
solar metallicities, Vernet et al. 2001; Humphrey et al. 2004 in prep.).  
We have calculated some basic parameters (fo\-llowing Heckman et
al. 1990) characterizing a superwind
which explains the kinematic properties and sizes of the quies\-cent giant halos:
(1)  energy injection rate $dE/dt$; (2)  dynamical time $t_{dyn}$; (3) infrared luminosity
expected from the starburst $L_{IR}$ ; (4) required star forming rate $SFR$; (5) 
supernova rate $SNR$; (6)
mass injection rate $dM/dt$. The results for all the objects are shown in Table 4.
We have assumed in our calculations a pre-shock density of 1 cm$^{-3}$ (Heckman
et al. 1990). Another parameter needed in the calculations is $v$, which is  half
the separation in velo\-city between the two components of the double peak profile
produced by the expanding bubble. The double peak is not resolved in our spectra
and we can only estimate an upper limit for $v$.
We estimate that two unresolved  emission lines separated by a given
velocity $\delta v$ would produce a line of FWHM$\geq$2$\delta v$ = 4$v$.
Therefore $v \leq \frac{FWHM}{4}$, where we have assumed FWHM = FWHM$_{min}$ 
in Table 2.
The values for $dE/dt$, $L_{IR}$, $SFR$, $SNR$ and $dM/dt$  are 
likely to be high upper limits (lower limits for $t_{dyn}$) due to the strong dependence
of these parameters with $v$ ($\propto v^3$). Typical $v$ values of
low redshift starburst driven galactic superwinds are seve\-ral hundred km s$^{-1}$
at several kpc from the center. $v$ values might be smaller for the LSBHs, since
the putative bubble has expanded to much larger distances (tens of kpc) and is likely to have
decelerated.

\begin{table*}
\centering
\begin{tabular}{cccccccccccc}   
\hline
	 (1)&	(2)	&	(3) &	(4) &	(5) &	(6) & (7) \\
Name & $dE/dt$	 &	$t_{dyn}$ &  log($L_{IR}/L_{\odot}$) & $SFR$ &	$SNR$ & $dM/dt$	 \\
    & $\times$10$^{45}$ erg s$^{-1}$ & $\times$10$^8$ yr & 	& M$_{\odot}$ yr$^{-1}$    & M$_{\odot}$ yr$^{-1}$ &M$_{\odot}$ yr$^{-1}$ &           \\ 
    &  $\leq$ &   $\geq$ &   $\leq$         &    $\leq$       & $\leq$   & $\leq$   \\ \hline
1809+407	& 0.9 	 & 3.2 	&  13.1	&  3450 & 26	&  	530	&  \\
0211-122	& 1.1 	 & 6.1	&  13.2	&  4200 & 32	&  	650	&  \\
1931+480	& 1.2 	 & 4.7	&  13.2	&  4550 & 34    & 	690	&  \\
0731+438	& 1.2 	 & 2.3	&  13.2	&  4570 & 35	&  	700	&   \\
2104-242	& 0.4 	 & 9.8	&  12.7	&  1370 & 11	&  	710	&  \\
1558-003	& 4.8 	 & 4.6	&  13.7	&  17970 & 138	&  	2760	&   \\
0828+193	& 0.6 	 & 9.2	&  12.9	&  2200 & 17	&  	340	& \\ 
0943-242	& 1.0	 & 5.9 	&  13.2 &  3830 & 29 	&  	590	&   \\ \hline 
1243+036 	& 1.0 (*)& 3.8 (*)&  13.2&  3000 (*) & 29	&  	255	&   \\ 
IRAS 00812-7112	& 0.6	&  	&  12.9	& 2000	& 	&	 300	\\ \hline
\end{tabular}
\caption{Outflows as the origin of the LSBHs. (2) Energy injection rate (3) dynamical time; 
(4) infrared luminosity expected from the starburst; (5) expected star 
forming rate; 
(6) supernova rate;
(7) mass injection rate. 
(*) Results from vO96. $\geq$ or $\leq$ indicate lower or upper limits
respectively. The values for IRAS 00812-7112 have been taken from 
Heckman et al. 1990.}
\end{table*}

The derived $L_{IR}$ upper limits are consistent with those measured
for ultraluminous infrared galaxies
(ULIRGs, $L_{IR}\geq$10$^{12}$L$_{\odot}$). Similar  rest frame FIR luminosities have also been 
estimated for high redshift radio galaxies from submm observations (Hughes, Dunlop \& Rawlings
 1997).

Since the calculated parameters are gross upper limits,
we can only say that a scenario such that the halos have
 been deposited by galactic winds 
is not excluded by the data set.

\subsection{Cosmological implications}

The discovery of the giant quiescent halos in all objects in the sample we have
studied suggests that
they are a common ingredient of high redshift radio galaxies. 
Although the nature of the halos -- be they disks, spherical envelopes,
 cooling flow
 nebulae, gas deposited by winds -- is not clear, they are  giant gaseous 
reservoirs within which the radio galaxy is embedded.  

Our results suggest that the halos surround the object completely.
The gas we see in emission  is the fraction of the halo inside the  ionization
cones which has been ionized by the quasar continuum. 
The gas outside these cones 
 is likely to have much lower ionization level.
Evidence for extended reservoirs of neutral gas 
associated with distant radio gala\-xies has been found by van Ojik et al. (1997;
 see also de Breuck et al. 2003).

We proposed in VM02 that the LSBHs could be part of the initial gas reservoir
from which the galaxy started to condense. 	According to galaxy
formation  models (e.g. Haiman, Spaans and Quataert 2000), dark matter halos collapse
in a cooling flow manner. As a result the baryons settle into rotationally
supported exponential disks embedded in the dark matter halo. Circular velocities
of $\sim$few hundred km s$^{-1}$ (consistent with our observations)
 would be expected for halos with masses $\ga$several $\times$ 10$^{12}$ M$_{\odot}$.

According to  Haiman, Spaans and Quataert (2000), 
the expected line widths are of order $\sim$30[(1+$z$)/6] \AA ~, which 
corresponds to $\sim$915 km s$^{-1}$ for HeII. The observed
FWHM of the lines from the quiescent halos can be easily reconciled with
the models taking into account that most disks are probably not seen edge on.

According to the models, due to Ly$\alpha$ cooling radiation,
the halos would have surface 
brightness $\sim$few $\times$ 10$^{-19}$ 
erg cm$^{-2}$ s$^{-1}$ arcsec$^{-2}$ out to the virial radius
($\geq$15 arcsec). However,
if a quasar switches on at some point and illuminates 
the halos, the expected surface
brightness would rise to 
$\sim$10$^{-16-17}$ erg cm$^{-2}$ s$^{-1}$ arcsec$^{-2}$
\cite{hai01} within the Str\"ogrem radius consistent with our observations.
The line widths, velocity shifts, measured angular sizes, line luminosities
 and surface brightness of the giant
quiescent halos are
 consistent with this scenario.

The interpretation of a cooling flow and a rotating disk do not
exclude each other, they could be two consecutive phases of the same process.
What is clear
in our case, is that the halos are not made of pristine material. They
 have been already enriched with heavy elements, which suggests
an advanced state in the process. We have also found
preliminary evidence for rotation in some of the
halos (although it is not possible to discard other interpretations).

	If this  galaxy formation scenario  is valid, 
we expect that similar 
structures exist associated with normal and active galaxies at different redshifts.
	The halos are likely to be more easily detected around powerful active
galaxies, since the active nucleus provides an intense supply of ionizing
photons.
In fact, giant quiescent (total extension $\geq$100 kpc) emission line 
halos  have been found in some $z\la$1 radio galaxies  
(e.g. Stockton, Ridgway \& Kellogg 1996, 
Sol\'orzano-I\~narrea, Tadhunter 
\& Axon 2001). 
 
	The quiescent LSBHs could be related (progenitors?) to the giant
HI disk like structures  found around some low redshift radio galaxies 
\cite{mor02} and elliptical galaxies \cite{oo02}. 
 An important difference is that the HI absorption line in
these objects is usually
much narrower ($\la$100 km s$^{-1}$) than the emission lines from 
the quiescent LSBHs. This suggests that the high redshift structures
are less  settled.

It is possible that similar giant gaseous structures 
have also been found around non active galaxies at different redshifts.
Steidel et al. (2000) discovered 
two giant  (physical extent $\geq$100 $h^{-1}$) diffuse
 Ly$\alpha$ emitters apparently associated with previously known Ly break galaxies
at $<z>=$3.09.
At low redshift, observations of  absorption line systems in the
spectra of background quasars have provided evidence for
large ($R\sim$100 $h^{-1}$ kpc) extended 
gaseous envelopes that surround galaxies
of a wide range of luminosity and morphological type (e.g. Chen, Lanzetta \& Webb 2001;
Lanzetta et al. 1995).  These galactic envelopes might  have similar origin and nature
as  the giant
quiescent halos we have discussed around high redshift radio galaxies.

	Therefore, we propose  a scenario such that a cooling flow has been triggered
as a result of  the collapse of a dark matter halo.  The baryons will end up settling
in a rotationally supported disk. During this process, the quasar has
switched on and illuminates and excites the gas. This
could happen before or after the gas has settled into the
 disk. The emission from the gas becomes easily detectable in this way, since its
surface brightness is much brighter than  expected from pure
  Ly$\alpha$ cooling radiation. During the process, intense star formation has already
taken place and outflows associated with supernovae explosions or the quasar activity 
have enriched the halos with heavy elements at tens of kpc from the nuclear region.

\section{Summary and conclusions}

	We have studied the  kinematic properties of the extended gas in 
a sample
of 10 high redshift ($z\sim$2.5) 
radio galaxies observed with the Keck II and VLT telescopes.
 We have discovered in all 10 HzRG (including 0828+193, VM02)
giant halos of quiescent gas   with 
FWHM(HeII)$\leq$850 km s$^{-1}$ and velocity shifts across the nebulae $\leq$600
 km s$^{-1}$. This component often
 extends over about 50 kpc  from the nuclear region and  sometimes beyond
the radio structures.
The emission line spectra of the quiescent halos are typical of active
galaxies.	 In addition, 8 out of 10 objects contain kinematically 
perturbed gas usually with high surface brightness and 
located inside the radio structures.

	We propose that we have isolated the emission
from  gas whose kinematics
has been perturbed by jet-induced shocks and the  
emission from  ambient non shocked gas (the quiescent halos). 

The quiescent halos often extend for more than 100 kpc and sometimes
beyond the radio structures.
	Typical Ly$\alpha$   surface brightness and luminosities
(within the slit)  are in
the range $\sim$few$\times$10$^{-17~to~-16}$ erg s$^{-1}$ cm$^{-2}$ arcsec$^{-2}$
and $\sim$10$^{43-44}$ erg s$^{-1}$. Estimated densities are in the range
$\sim$17-150 cm$^{-3}$.
The quiescent halos  
are enriched with heavy ele\-ments at tens of kpc from the  active nucleus.
The  quasar
continuum is the dominant excitation mechanism of the quies\-cent halos
along the radio axis.  
The implied quasar 
luminosities are in the range $\sim$several$\times$10$^{45}$-10$^{47}$
 erg s$^{-1}$,
in the same range as radio loud quasars at comparable redshift.
 Cooling radiation,
shock photoionization and stellar photoionization
 cannot explain at the same time all 
 measured properties (line luminosities, surface brightness, 
sizes, line ratios) of the quiescent halos along the radio axis. 
We do not reject the possibility that other mechanisms  excite
 the gas  in other regions  of
the halos.

	The quiescent halos are likely to be  giant gaseous reservoirs
(rather than companion galactic satellites).   
The detection  in all objects suggests that
they could be a common ingredient of high redshift radio galaxies. 
Whatever their nature (rotating disks, spherical envelopes, 
cooling flow nebulae, gas deposited by galactic winds)
the radio galaxy seems to be embedded within the halos.

	We propose a scenario such that the quiescent halos are part of dark matter
halos that collapsed in  the early phases of formation of what will become a 
giant elliptical galaxy.
A cooling flow is generated.  For the more evolved halos,
 the baryons have  settled in a rotationally
supported disk. In other objects, the cooling flow is  in an earlier phase.
At some point, a quasar has switched on and the halos have become 
observable thanks
to strong line emission powered by the quasar continuum.
Intense star formation has already been triggered and associated
 stellar winds 
 or quasar
outflows have
enriched the halos with heavy elements at tens of kpc from the nucleus. 
 The measured sizes, emission line widths, velocity shifts, presence of heavy elements, 
surface
brightness 
and emission line luminosities of the quiescent halos 
are consistent with this  scenario. 
2D spectroscopy is now critical to map the kinematic, ionization and
morphological properties of the halos in two spatial dimensions.

\section*{Acknowledgments}
 We thank  Marshall Cohen
for his help to collect the Keck  data and
useful scientific discussions. 
We thank Andrea Cimatti and Bob Goodrich
 for  their support to collect the Keck set of data.
We thank 
 Chris Carilli for providing the radio VLA maps 
and the Leiden HzRG group for the 2104-242 VLT spectrum.
RAEF is affiliated to the Astrophysics Division, Space Science Department, 
European Space Agency.

\vspace{0.3cm}
\begin{figure*}
\includegraphics{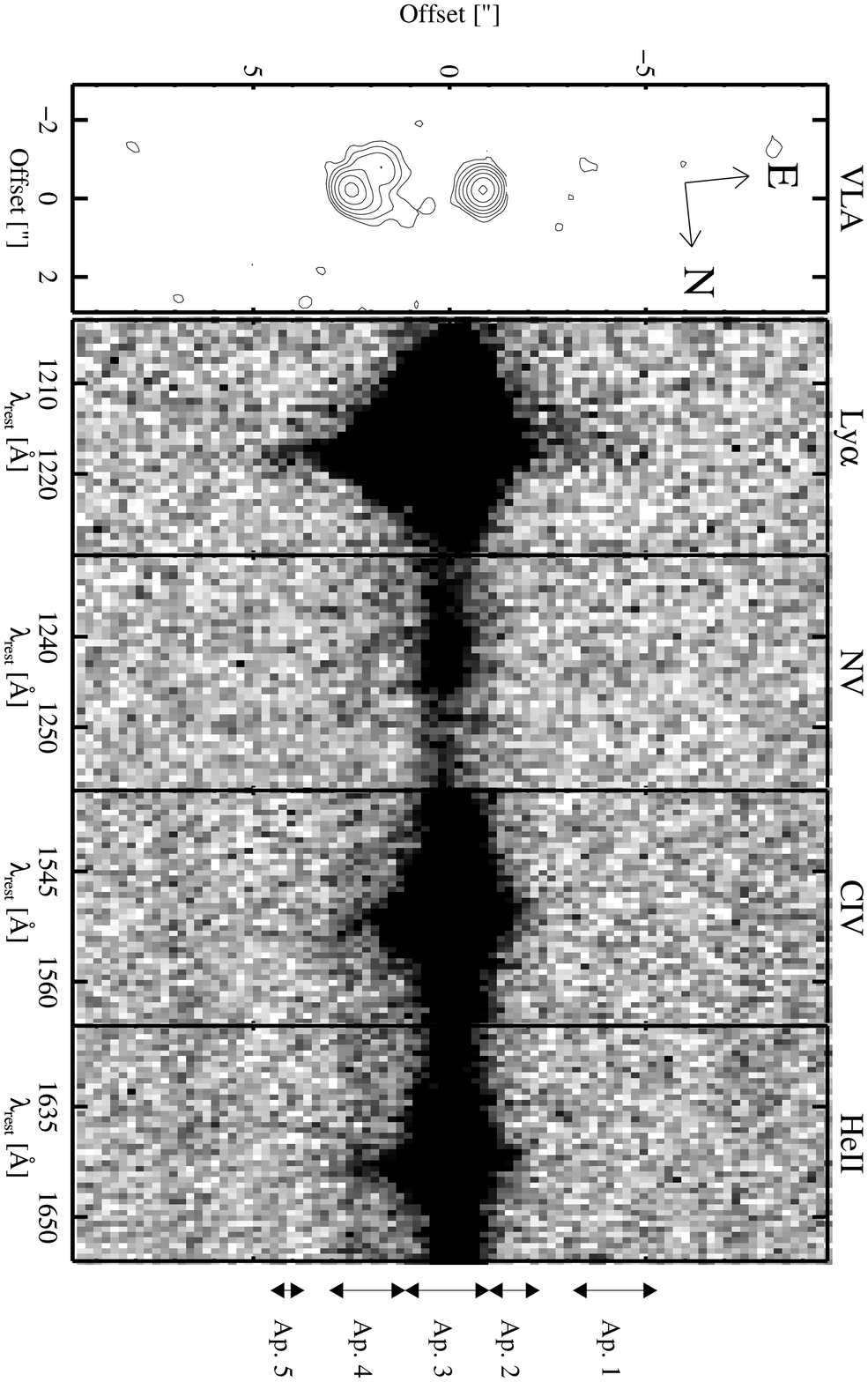}
\includegraphics{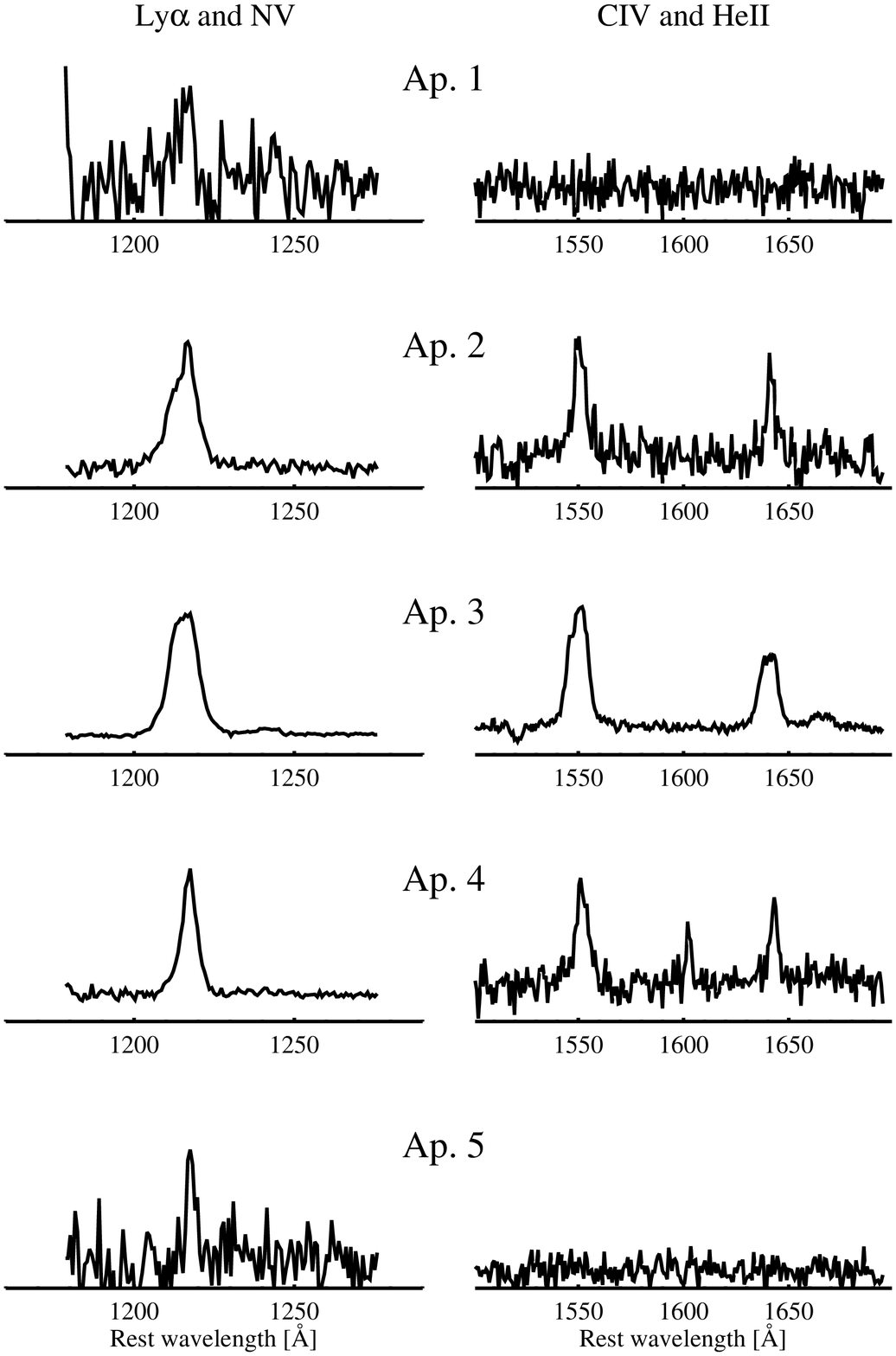}
\includegraphics{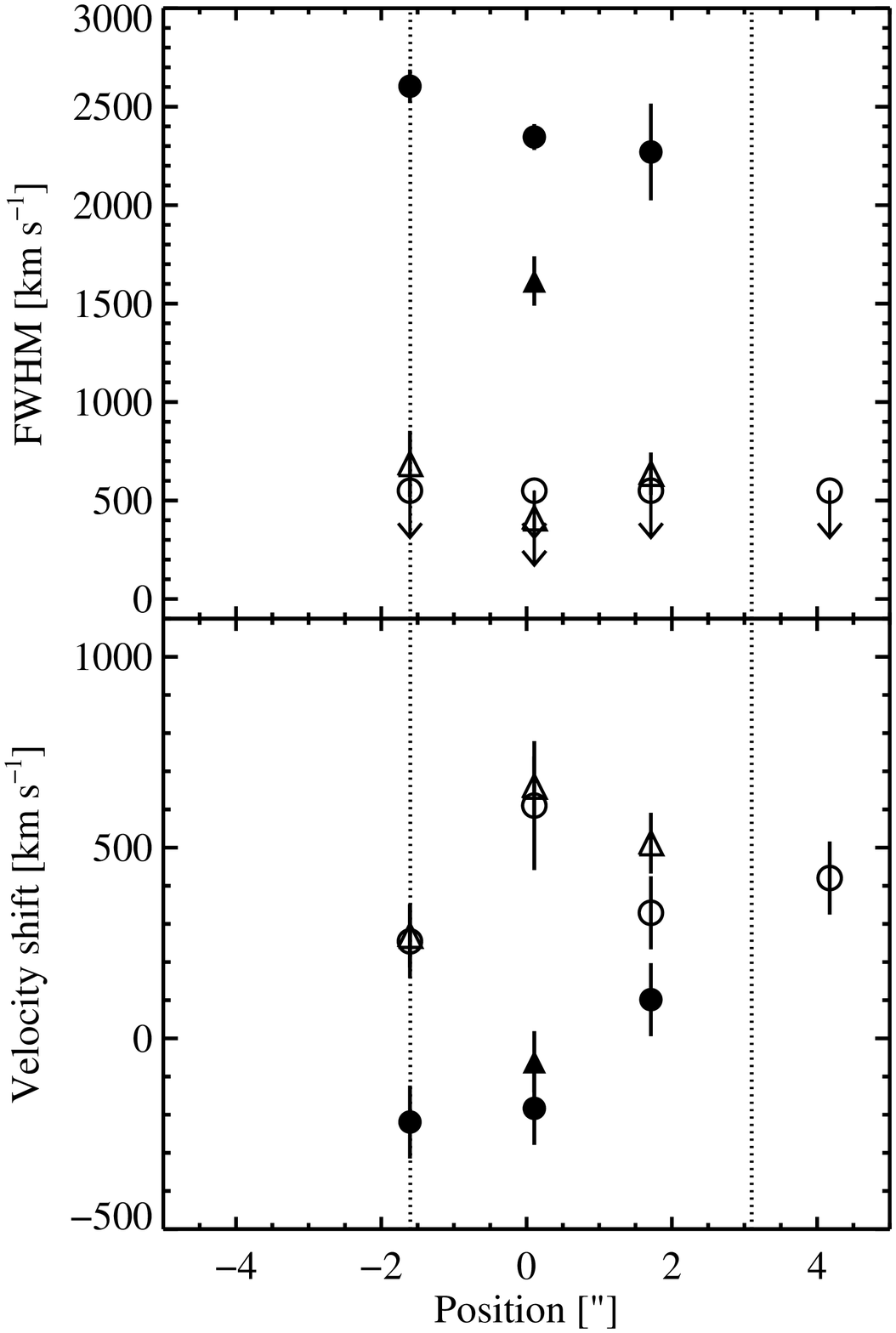}
\vspace{8in}
\caption{{\bf 1809+407:} {\bf Top}: VLA radio
 map spatially aligned with the 2D
Keck spectra of the main emission lines.  The apertures selected for the kinematic
analysis are also shown. The spatial zero in all 
figures is the position of the continuum centroid measured on
the Keck spectra.  Notice the LSBH extending beyond the radio structures with 
quieter   kinematics
compared to the high surface brightness regions.
{\bf Bottom left}: Spatially integrated spectra for the selected apertures. 
The flux scales are arbitrary to show clearly the emission
lines. {\bf Bottom right}: Kinematic properties of the ionized gas. 
FWHM  and velocity shift  
 relative to the HeII emission at the continuum
centroid are shown for the two 
kinematic components 
revealed
by the spectral fit to  Ly$\alpha$ and HeII.  
Circles: Ly$\alpha$; triangles: HeII. Open symbols: narrow
component; solid symbols: broad component. Arrows represent upper limits. The vertical lines mark
the outer edge of the radio source. 
The narrow component has FWHM in the range $\sim$500-700 km s$^{-1}$
and maximum velocity shift of 400 km s$^{-1}$ across the nebula
(as measured for HeII). The results are in good agreement with those
obtained for Ly$\alpha$.}
\end{figure*}

\begin{figure*}
\includegraphics{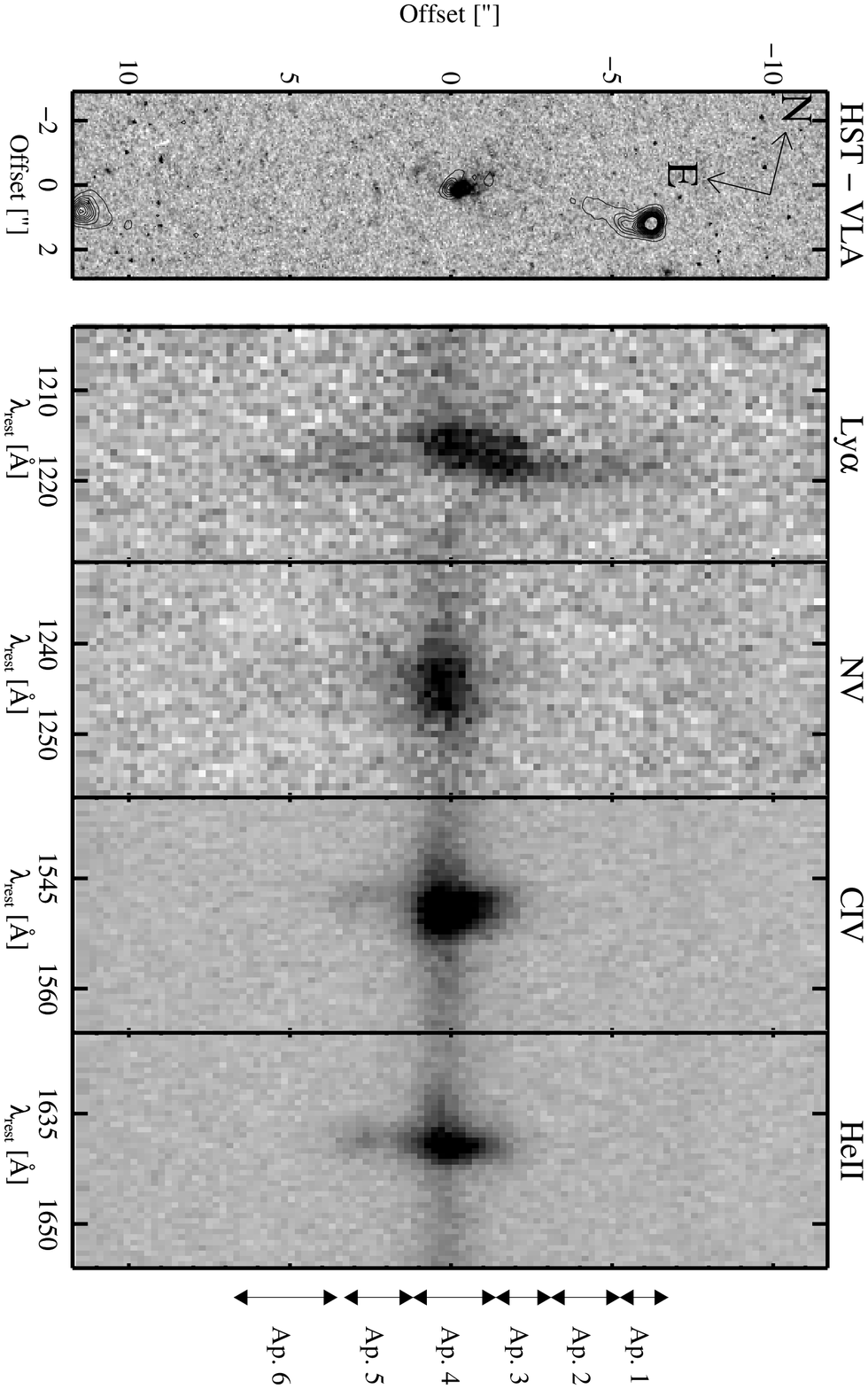}
\includegraphics{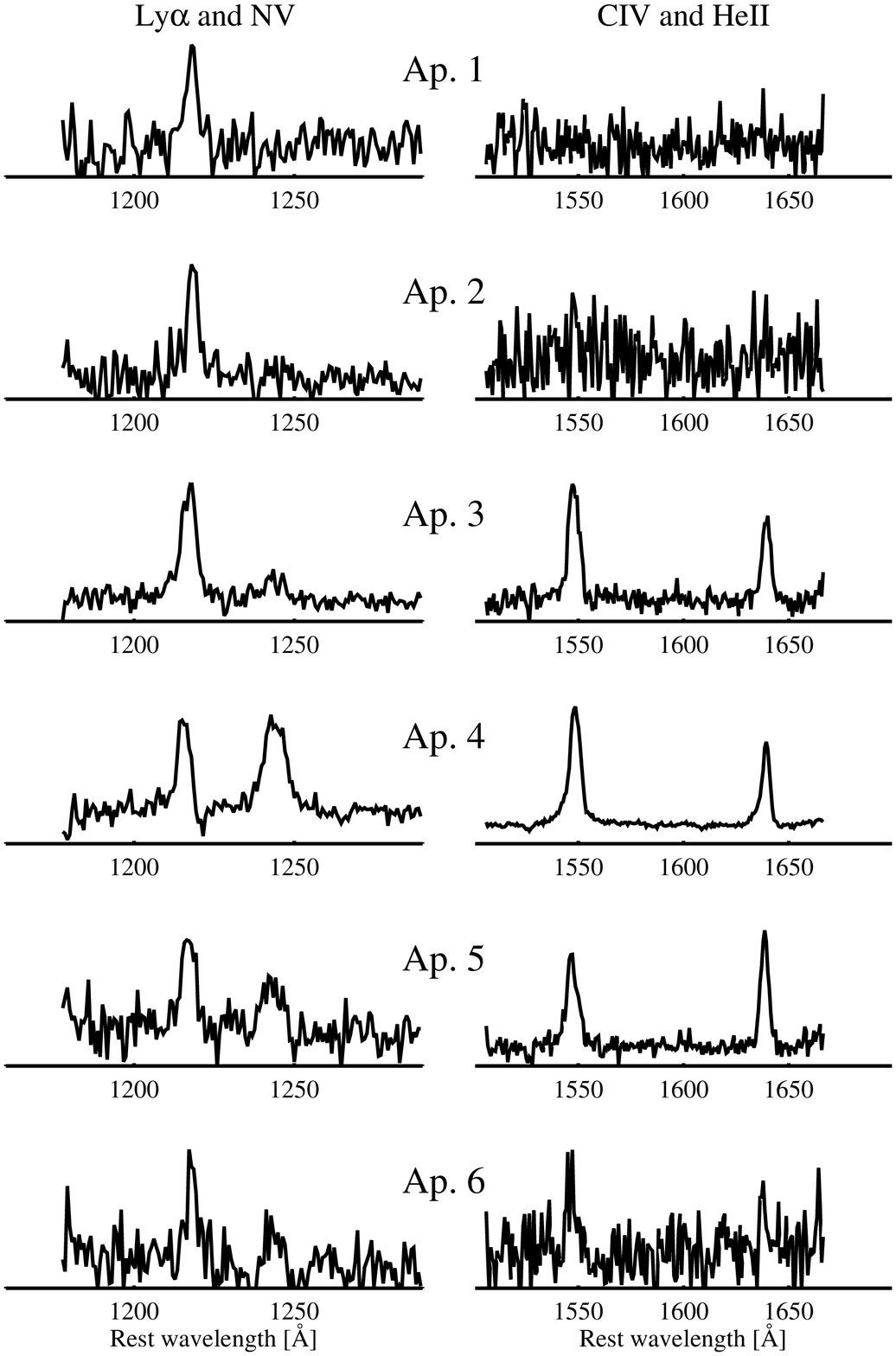}
\includegraphics{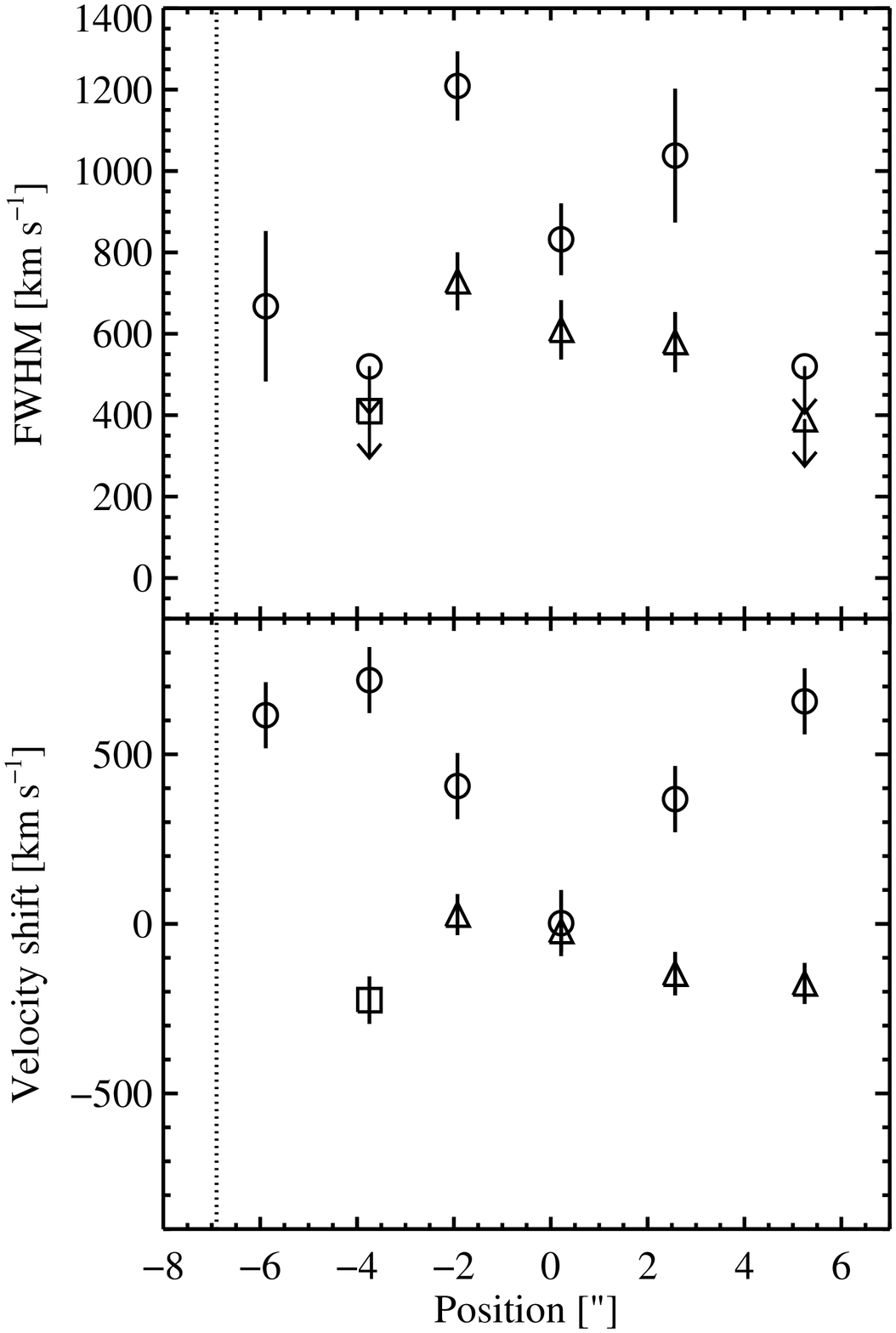}
\vspace{8.5in}
\caption{{\bf 0211-122:} Symbols and lines as in Fig.~1.
A very extended LSBH 
	is detected in the 2D
spectra of all emission lines (top).  In all figures
results for CIV  are only  shown 
when they provide
convincing evidence for the presence of low velocity gas, a signature of the
quiescent halos.   Both CIV
(squares) and HeII are very narrow in the outer regions of the object
(FWHM$\leq$400 km s$^{-1}$, bottom right panels). HeII has FWHM in the range $\leq$400-700 km s$^{-1}$
and maximum velocity shift of 200 km s$^{-1}$ across the nebula. 
Ly$\alpha$ is heavily
absorbed   at some spatial positions  and the kinematic
results obtained with this line are not reliable, especially in the
inner apertures.
Spectral decomposition of the lines was not performed for this object.
}
\end{figure*}

\begin{figure*}
\includegraphics{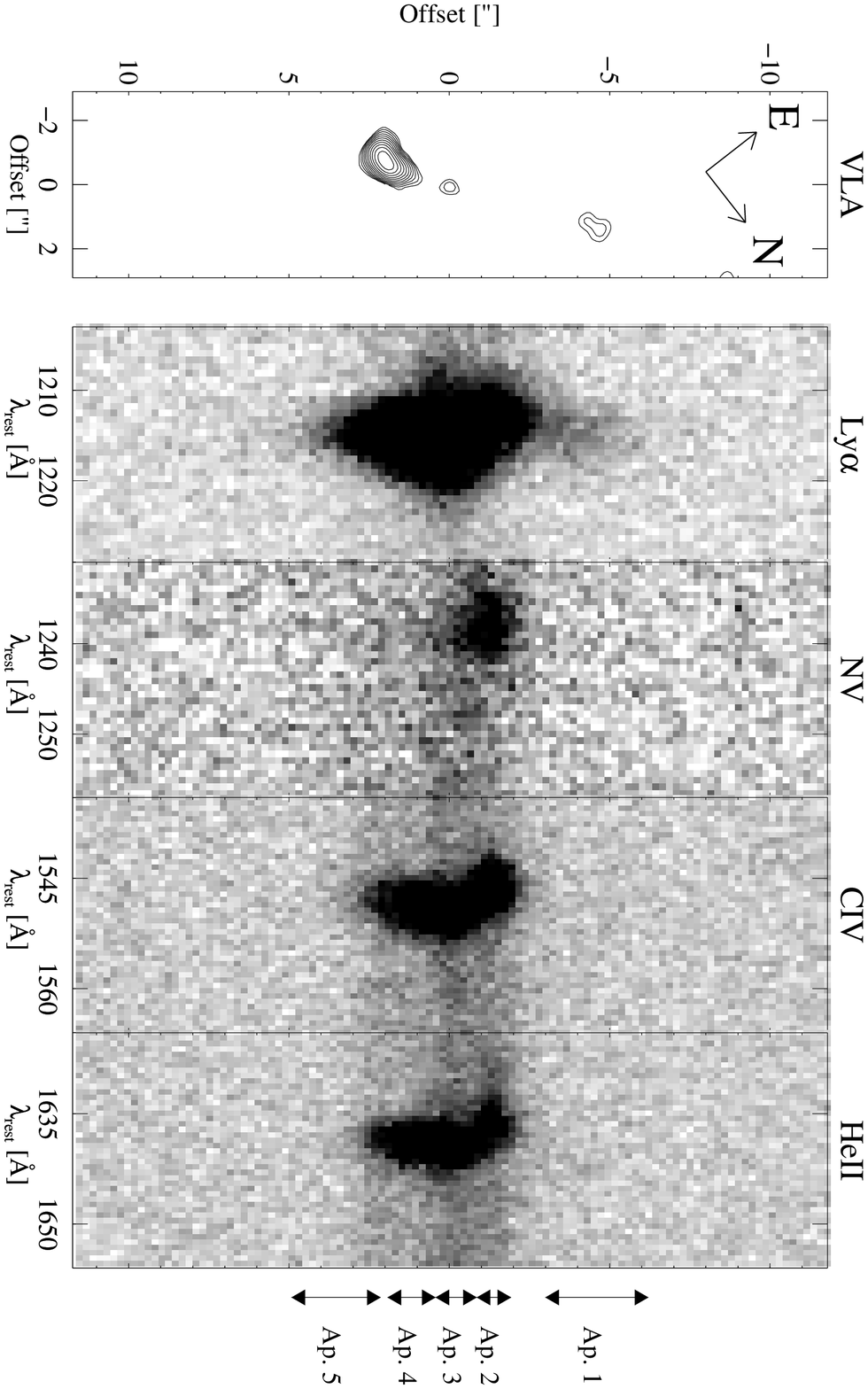}
\includegraphics{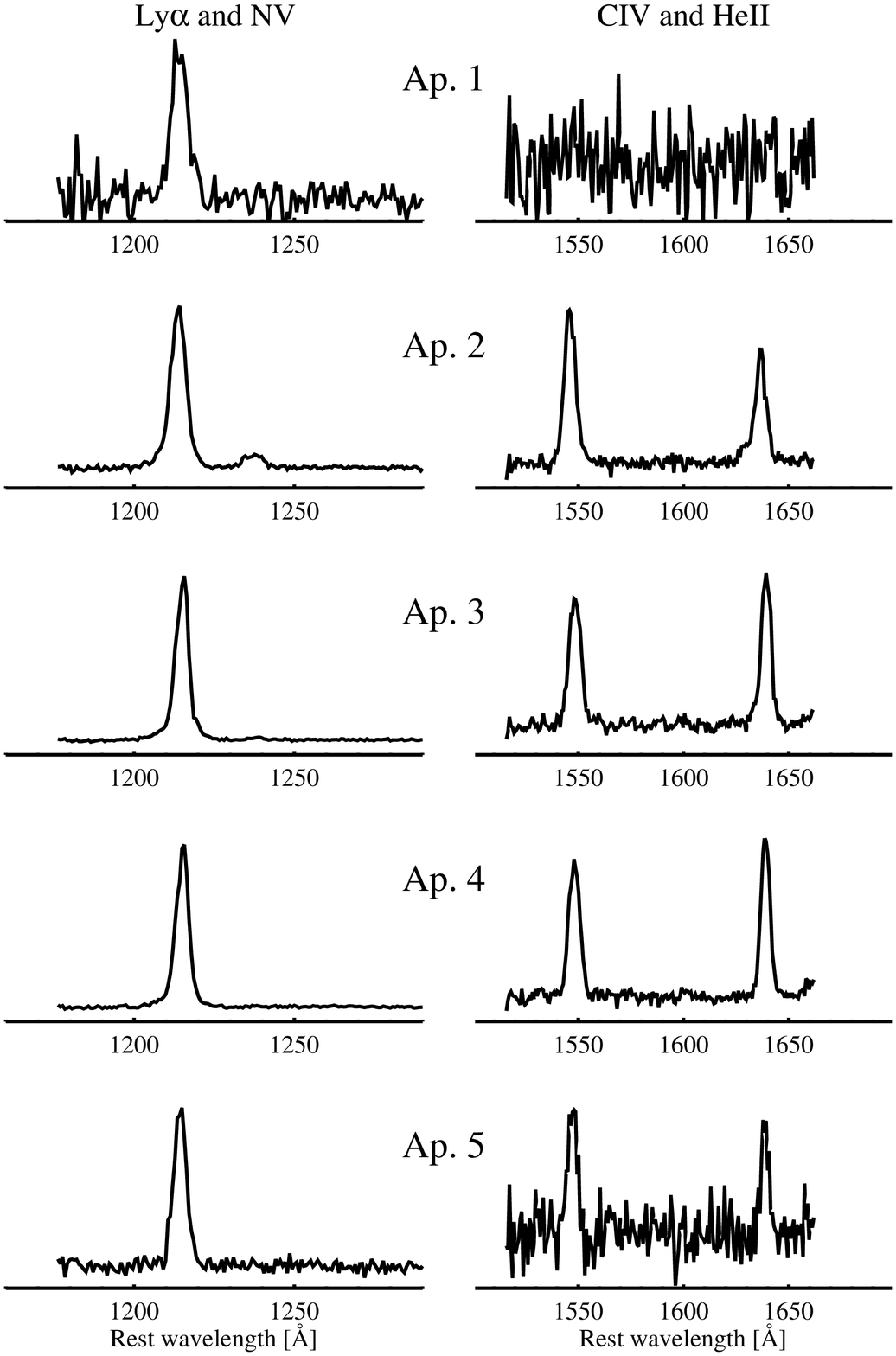}
\includegraphics{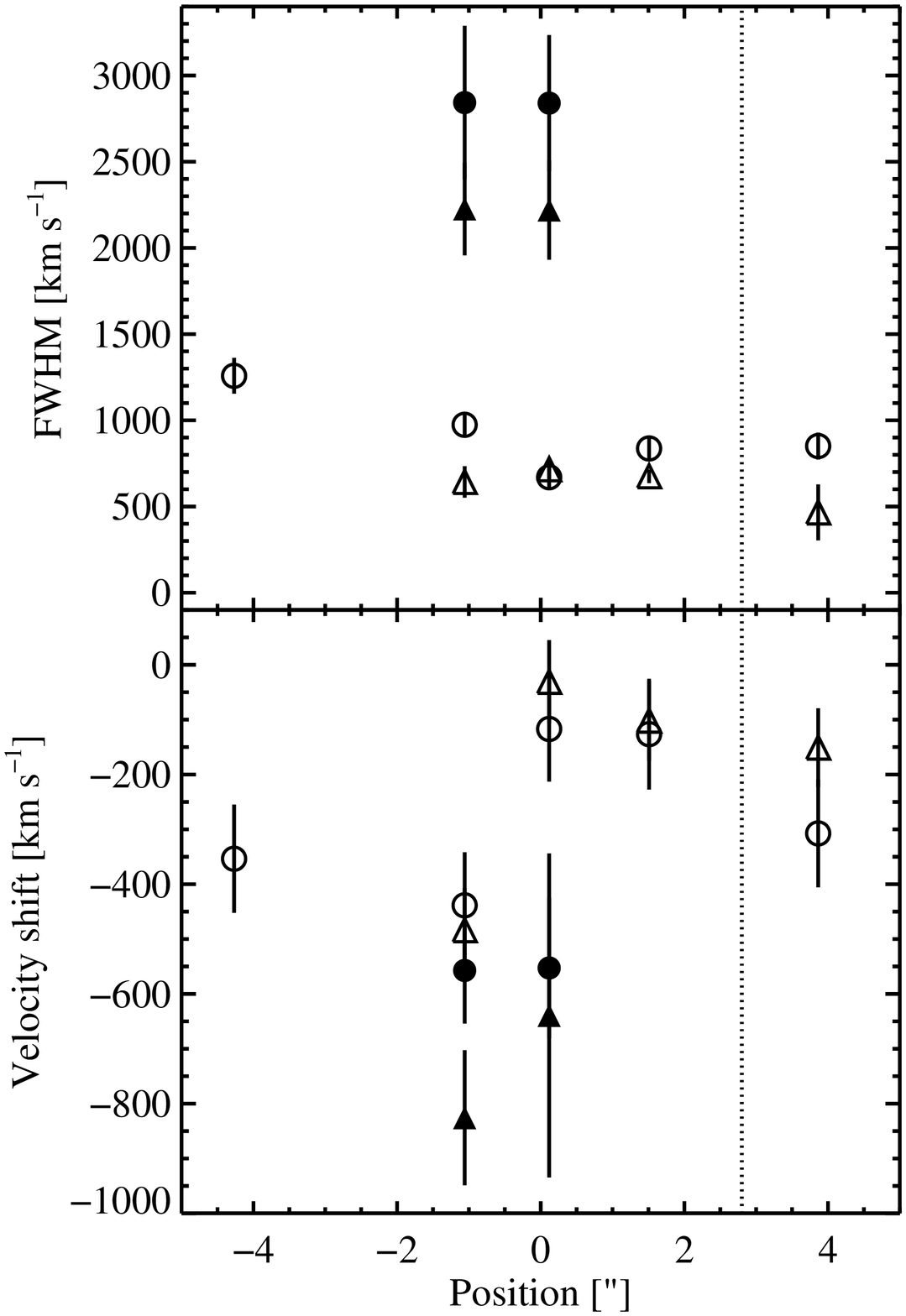}
\vspace{9in}
\caption{{\bf 1931+480}: Symbols and lines as in Fig.~1. A LSBH with 
apparently more quiescent kinematics compared to the high surface
brightness regions is detected in the 2D Ly$\alpha$ spectrum (top).
The fits to the HeII line reveal (bottom right panels)
a  narrow component with FWHM in the range $\sim$450-720 km s$^{-1}$
and maximum velocity shift of 450 km s$^{-1}$ across the nebula. 
Similar results are obtained for
 Ly$\alpha$.}
\end{figure*}

\begin{figure*}
\includegraphics{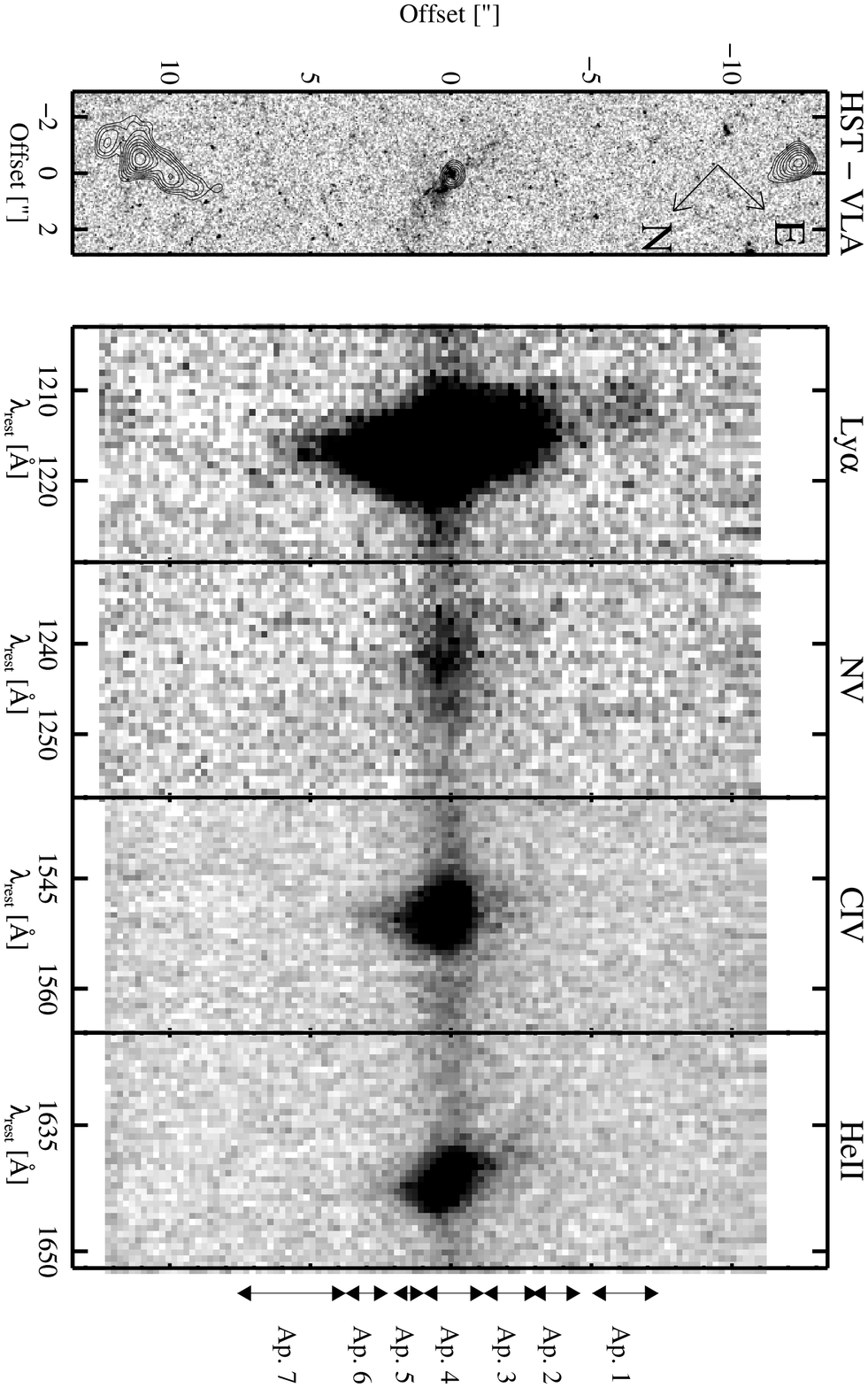}
\includegraphics{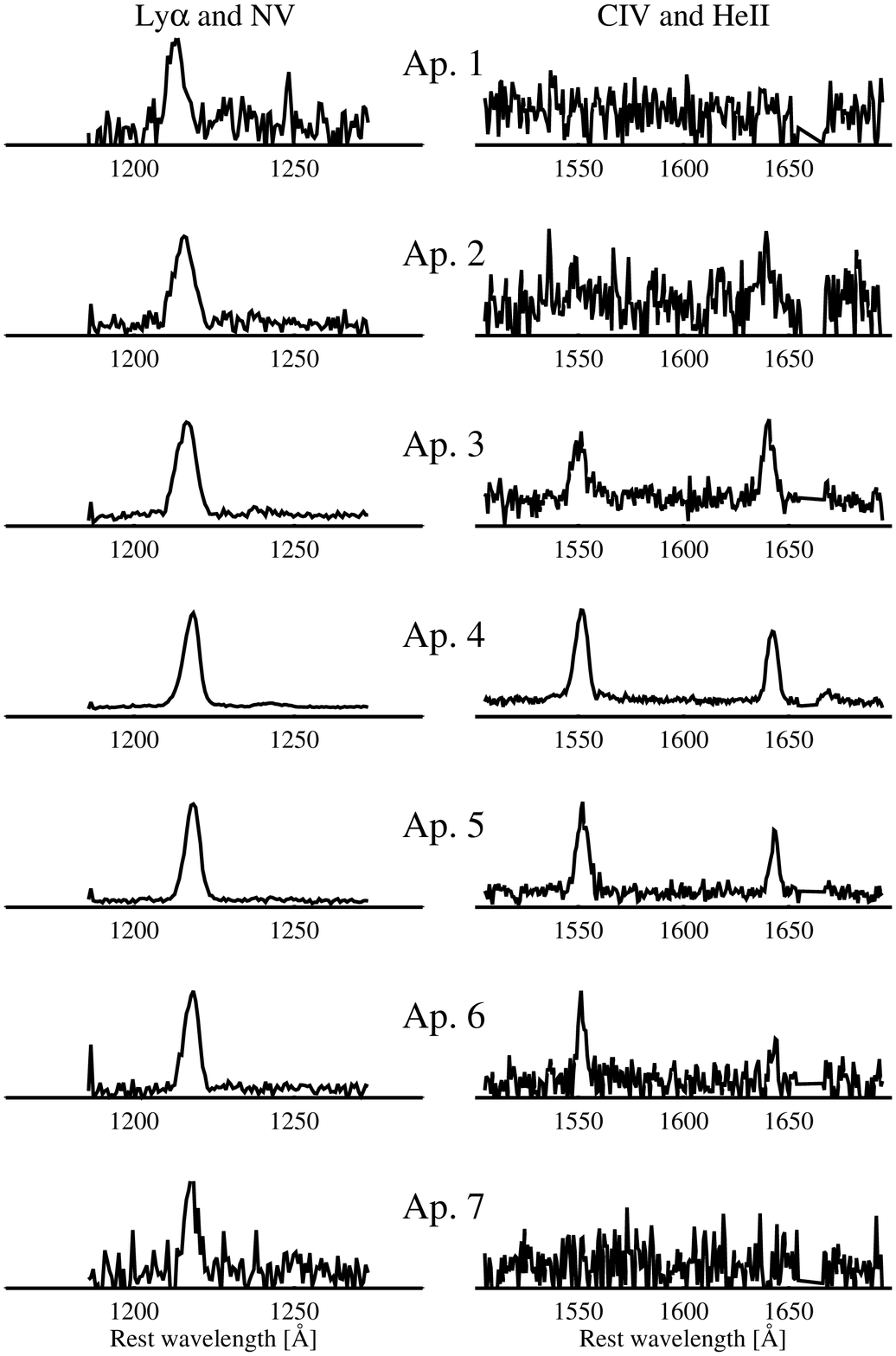}
\includegraphics{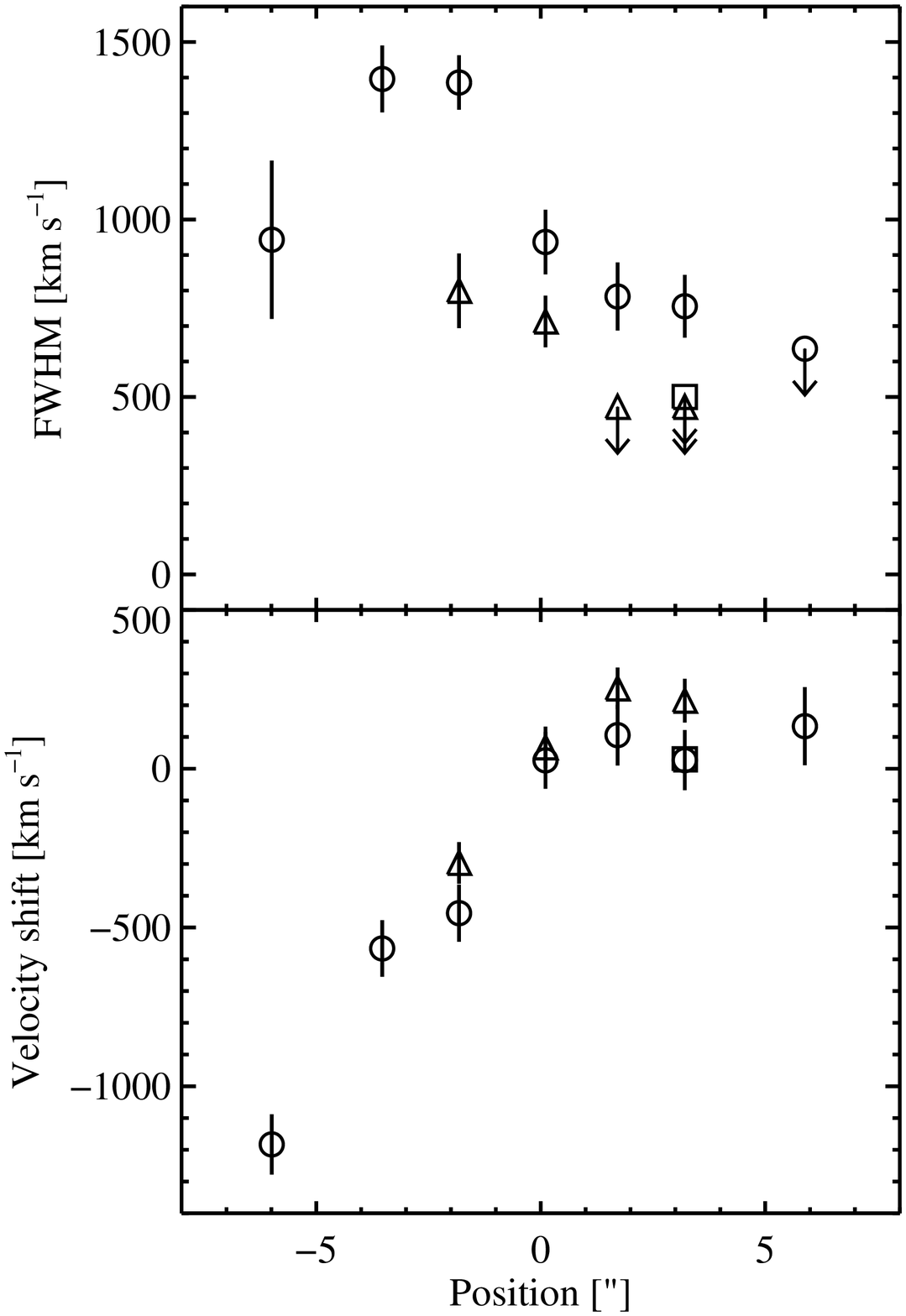}
\vspace{8.5in}
\caption{{\bf 1410-001:} Symbols and lines as in Fig.~1.
An extended LSBH 
	is detected in the 2D
line spectra (top panels) with apparently more quiescent kinematics than 
the high surface brightness regions. The fits to HeII, Ly$\alpha$ and CIV
(bottom right panels) confirm the existence of narrow 
line emitting gas across several
apertures with FWHM(HeII) in the range $\leq$472-800 km s$^{-1}$ and 
maximum velocity shift $\sim$500 km s$^{-1}$.}

\end{figure*}
\begin{figure*}
\includegraphics{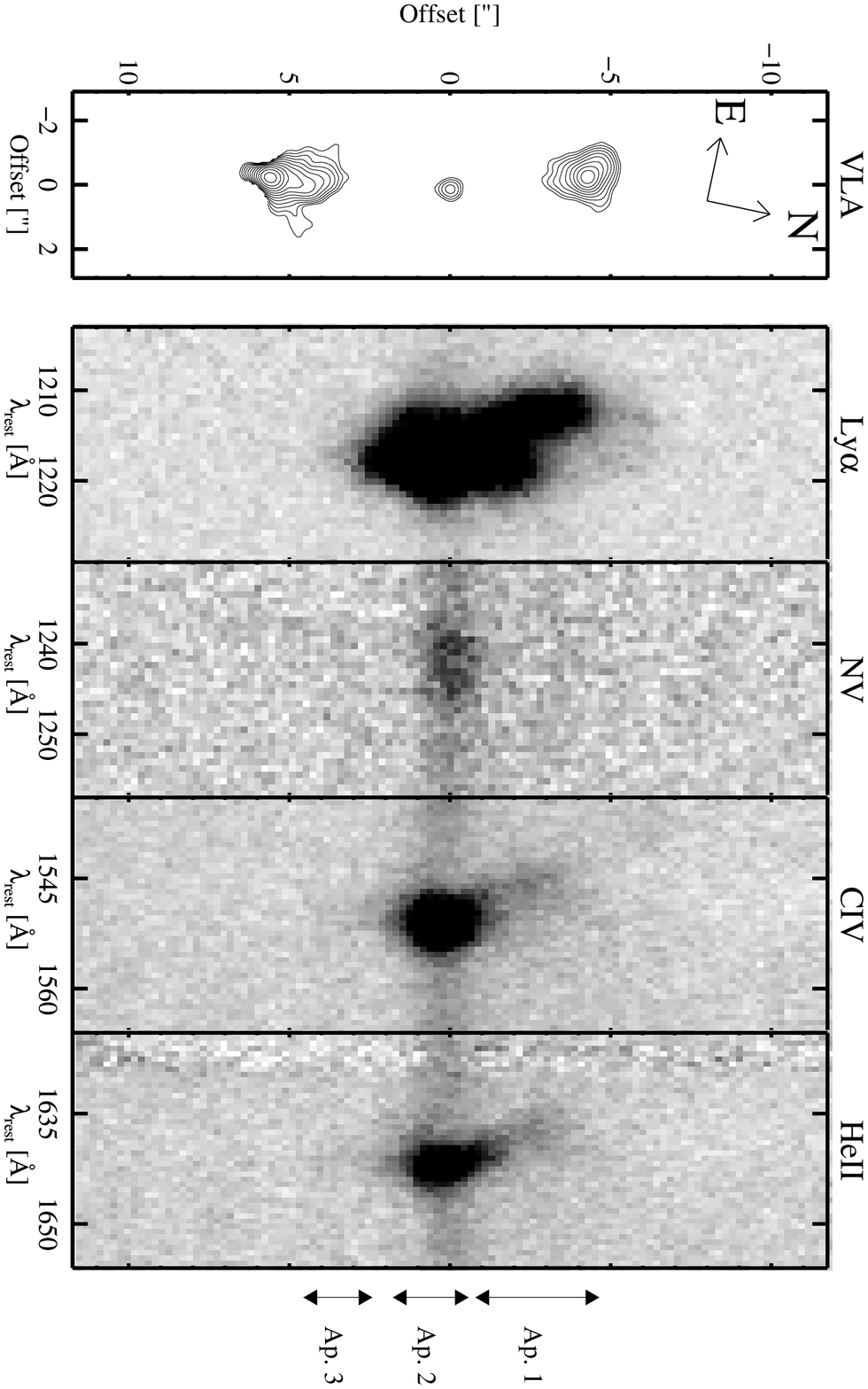}
\includegraphics{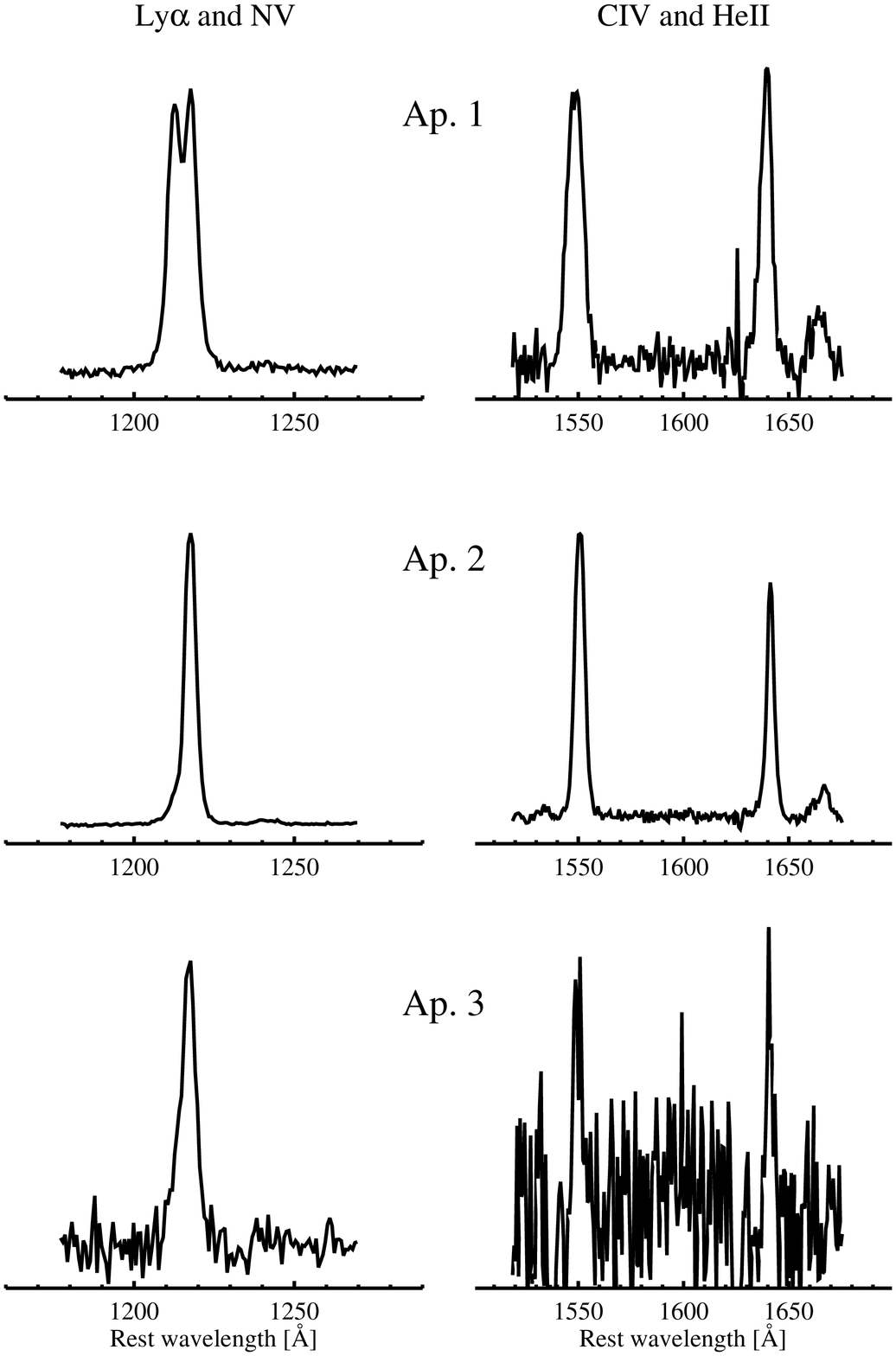}
\includegraphics{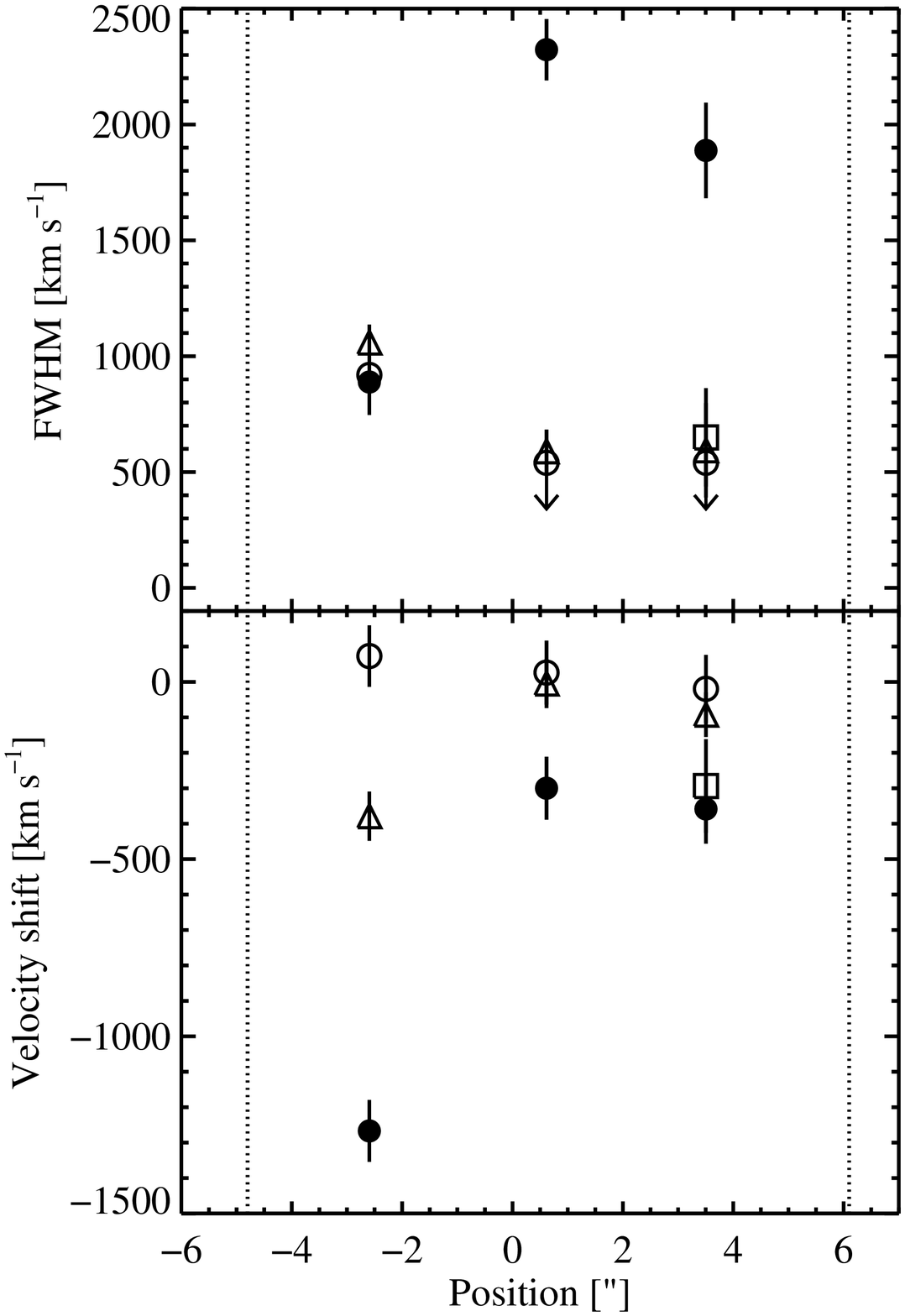}
\vspace{9in}
\caption{{\bf B3 0731+438:} Symbols and lines as in Fig. 1.
The double Ly$\alpha$ peak in the spectrum of ap. 1 (bottom left
panels)  is due to the contribution of the separate component
clearly seen in the 2D spectrum (top). We have not considered this
as part of the quiescent LSBH. A narrow
component is found across the rest of the object, 
with FWHM$\sim$600 km s$^{-1}$ and velocity 
shift$\leq$100 km s$^{-1}$ (bottom right panel). The results obtained with the three
lines are in good agreement.}
\end{figure*}

\begin{figure*}
\includegraphics{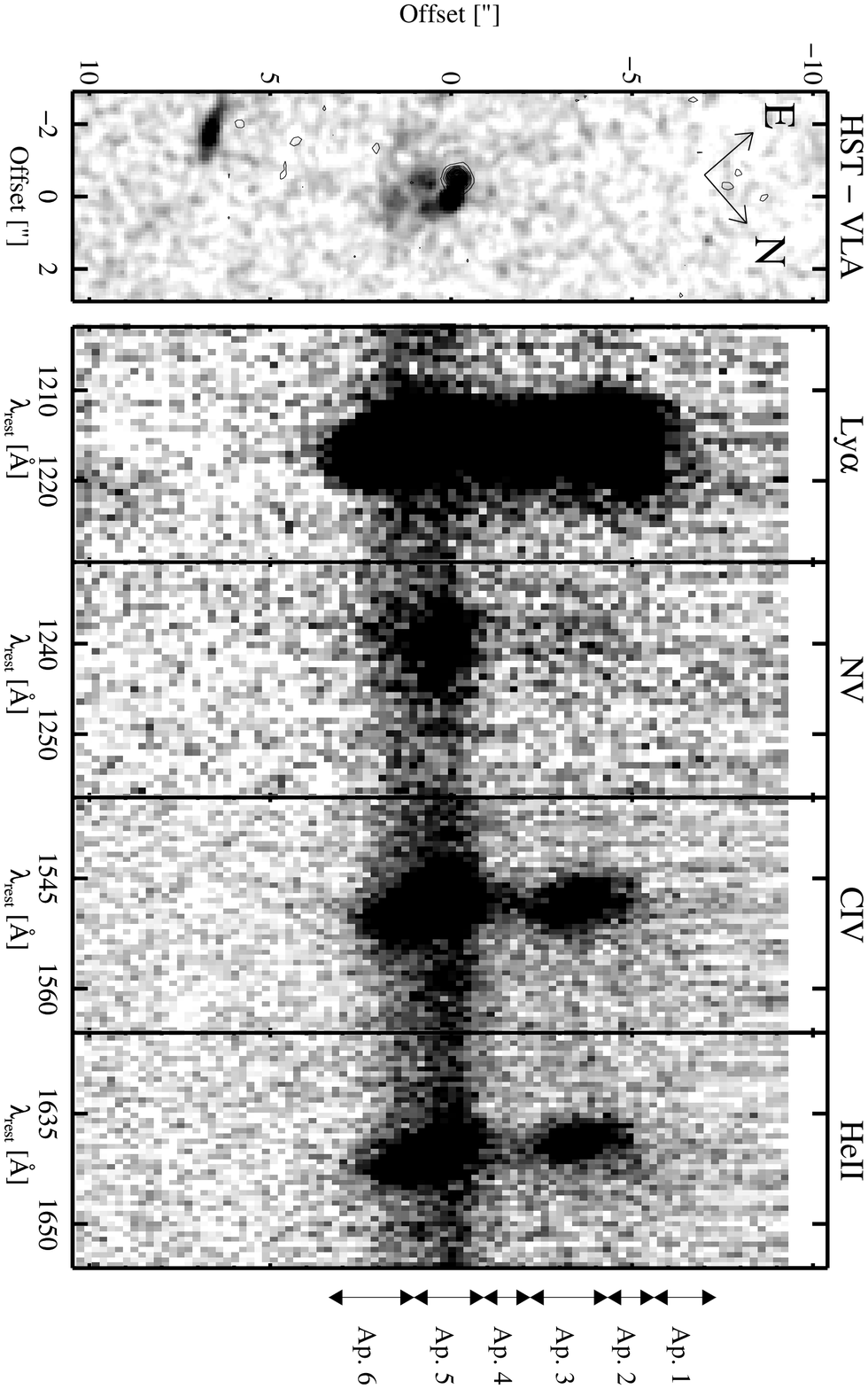}
\includegraphics{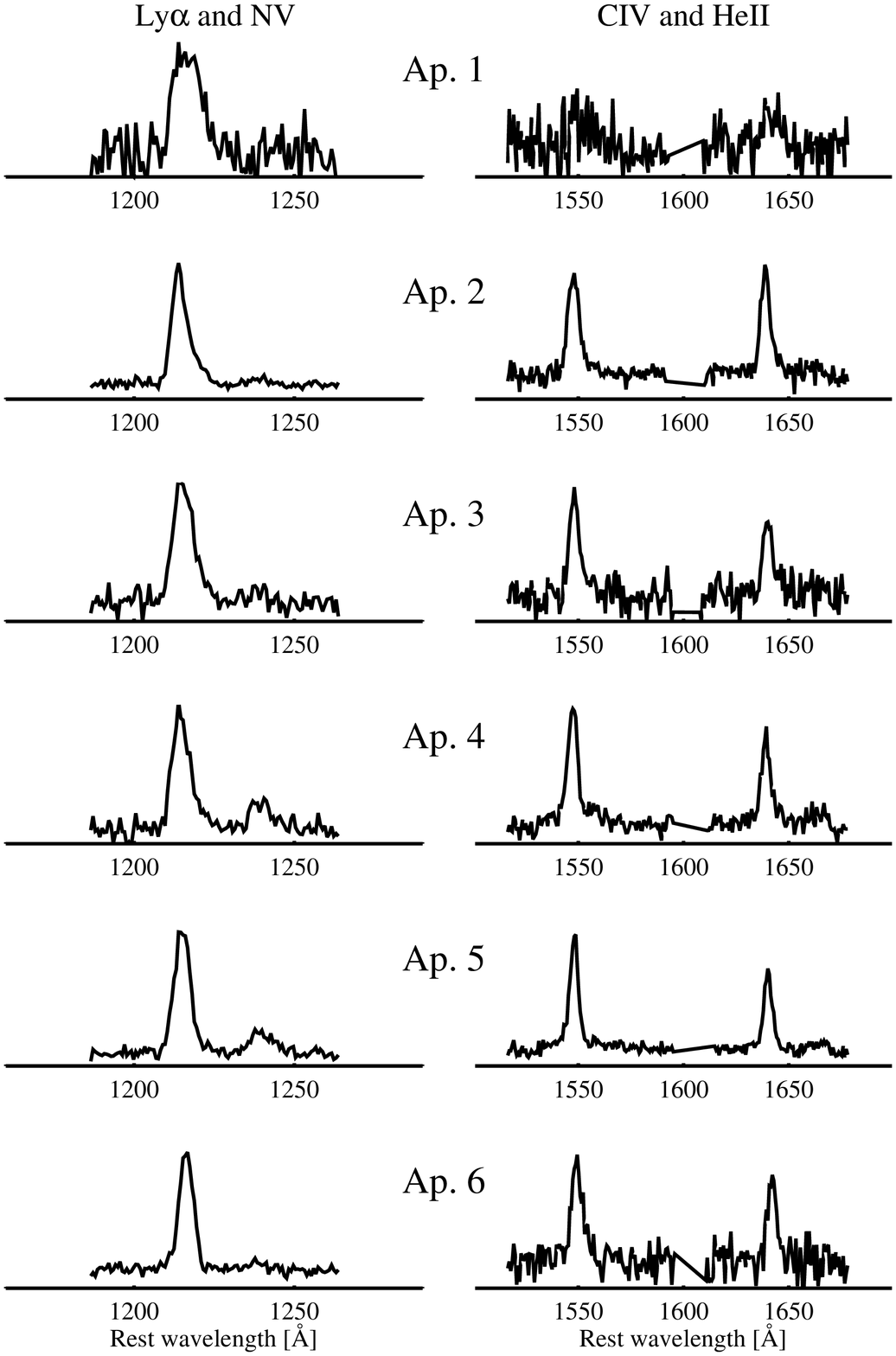}
\includegraphics{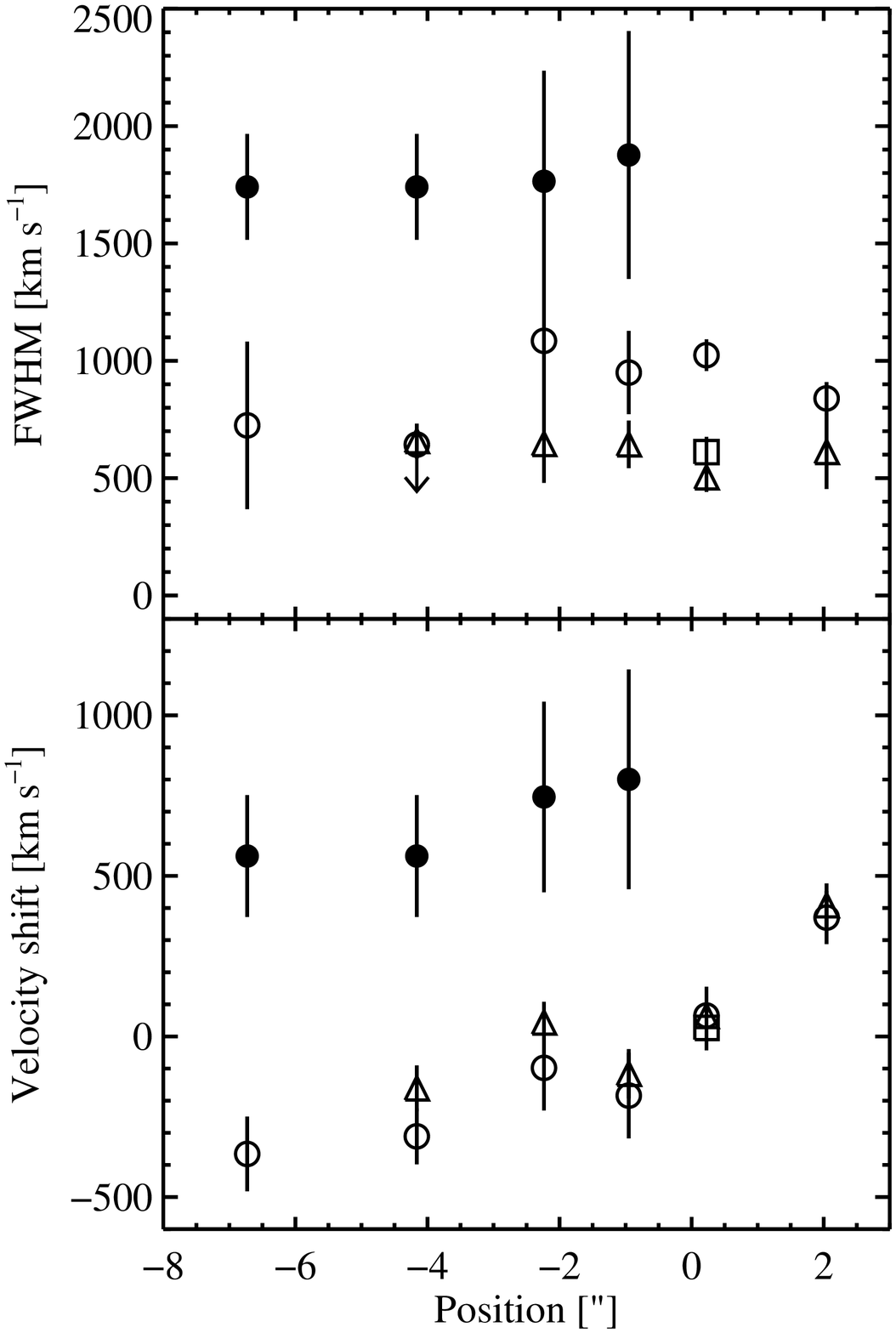}
\vspace{9in}
\caption{{\bf 2105+236:} Symbols and lines as in Fig.~1.
Contrary to other objects, there
 is no evidence for a quiescent halo in the 2D line spectra
of this object. However, the kinematic analysis (bottom right panels)
reveals the
presence of a narrow component with FWHM(HeII)$\sim$500-650 km s$^{-1}$
and maximum velocity shift $\sim$570 km s$^{-1}$.
}
\end{figure*}

\begin{figure*}
\includegraphics{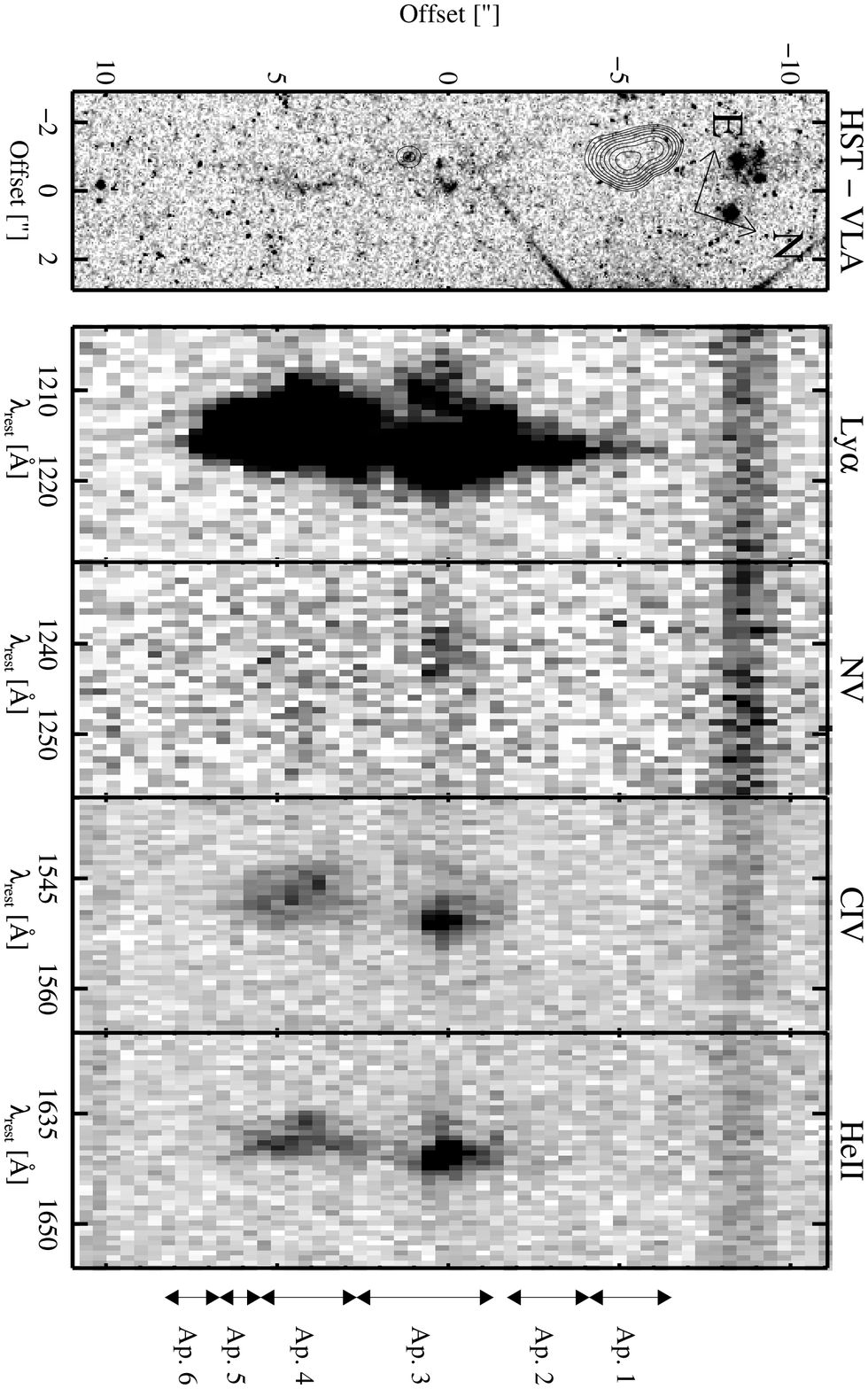}
\includegraphics{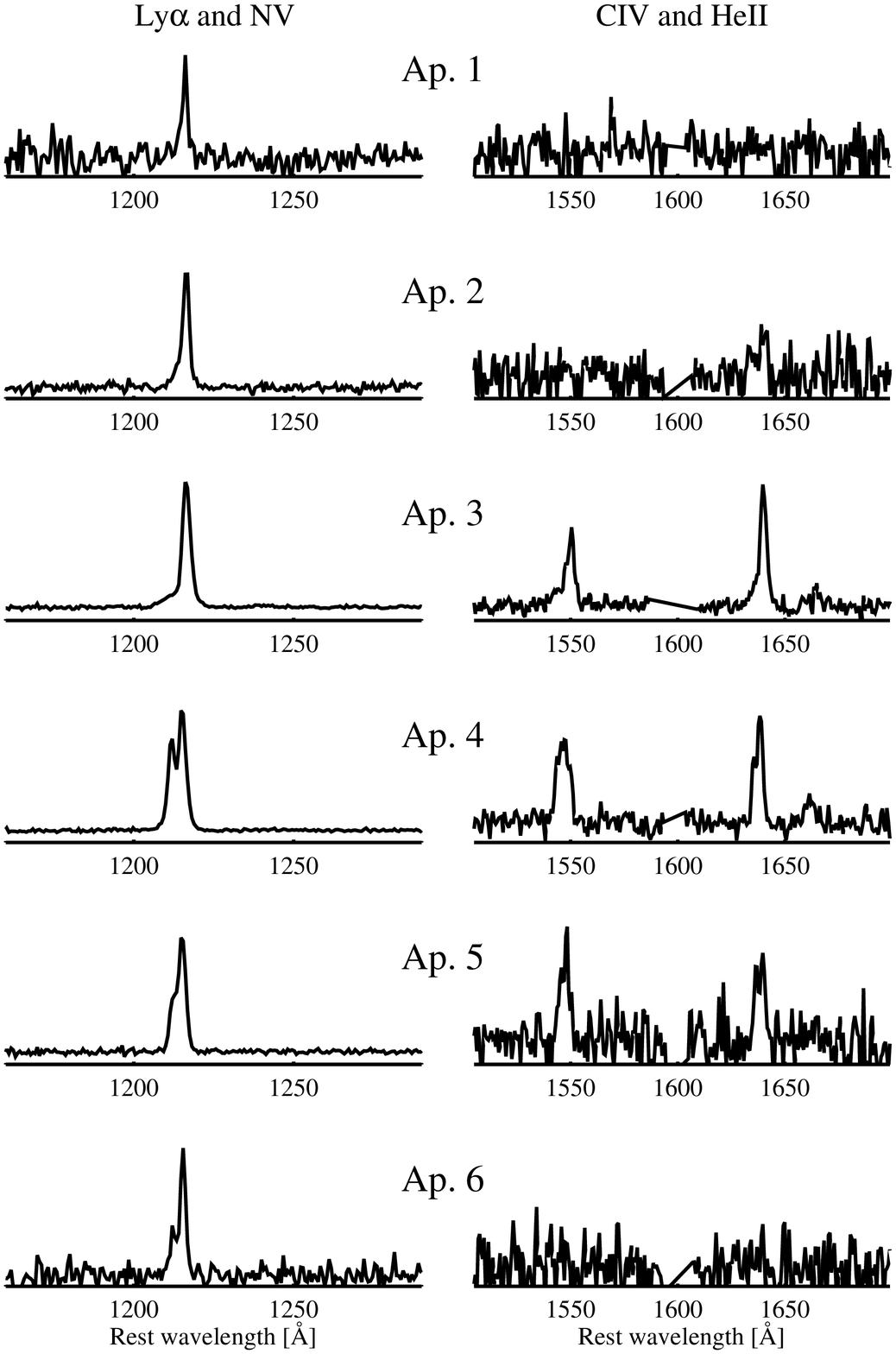}
\includegraphics{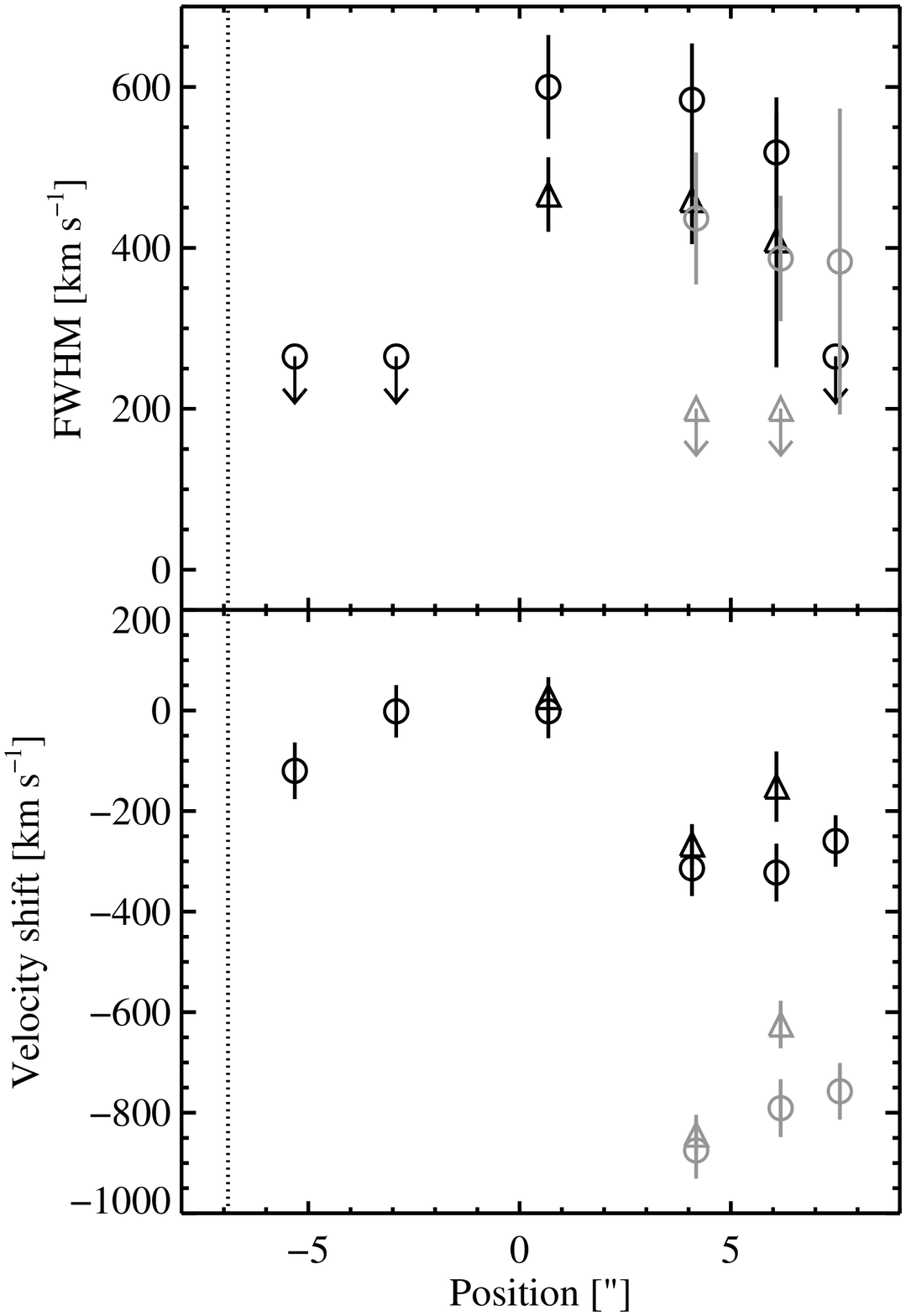}
\vspace{8.5in}
\caption{{\bf 2104-242:} 
The spectra for this object were obtained at higher resolution (see
text). Particularly interesting is the very narrow (FWHM$\leq$265 km s$^{-1}$)
Ly$\alpha$ emission detected in the outer regions of the object (bottom right panels), 
clearly seen extending 
to the NE in the 2D Ly$\alpha$ spectrum. This is the gas we have considered
as the quiescent LSBH.
Two kinematic components are found in ap. 3 to 5.
We have represented them with black and grey colours (rather than solid and
open symbols) to highlight the fact that no broad components are found in this
object.}
\end{figure*}

\begin{figure*}
\includegraphics{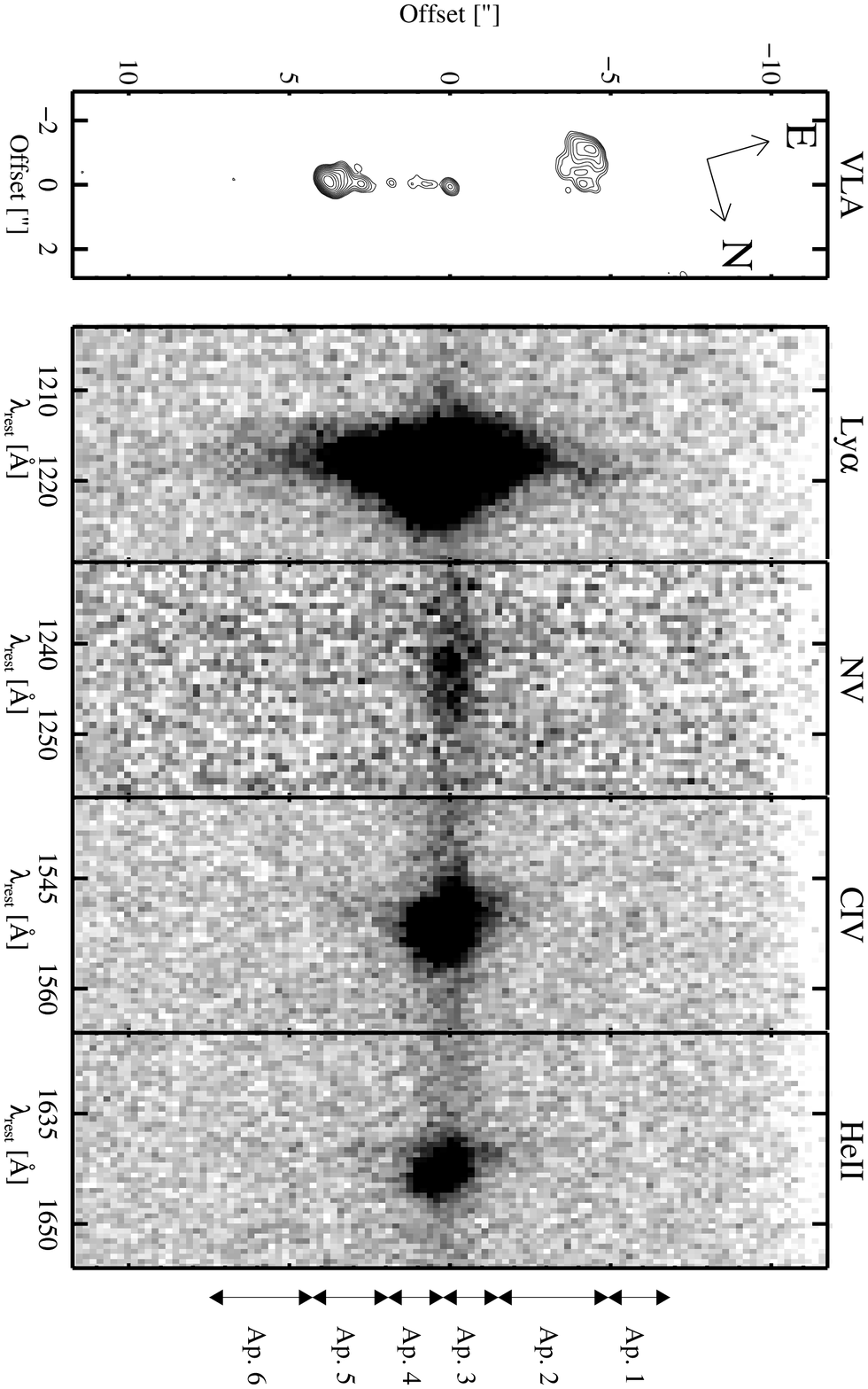}
\includegraphics{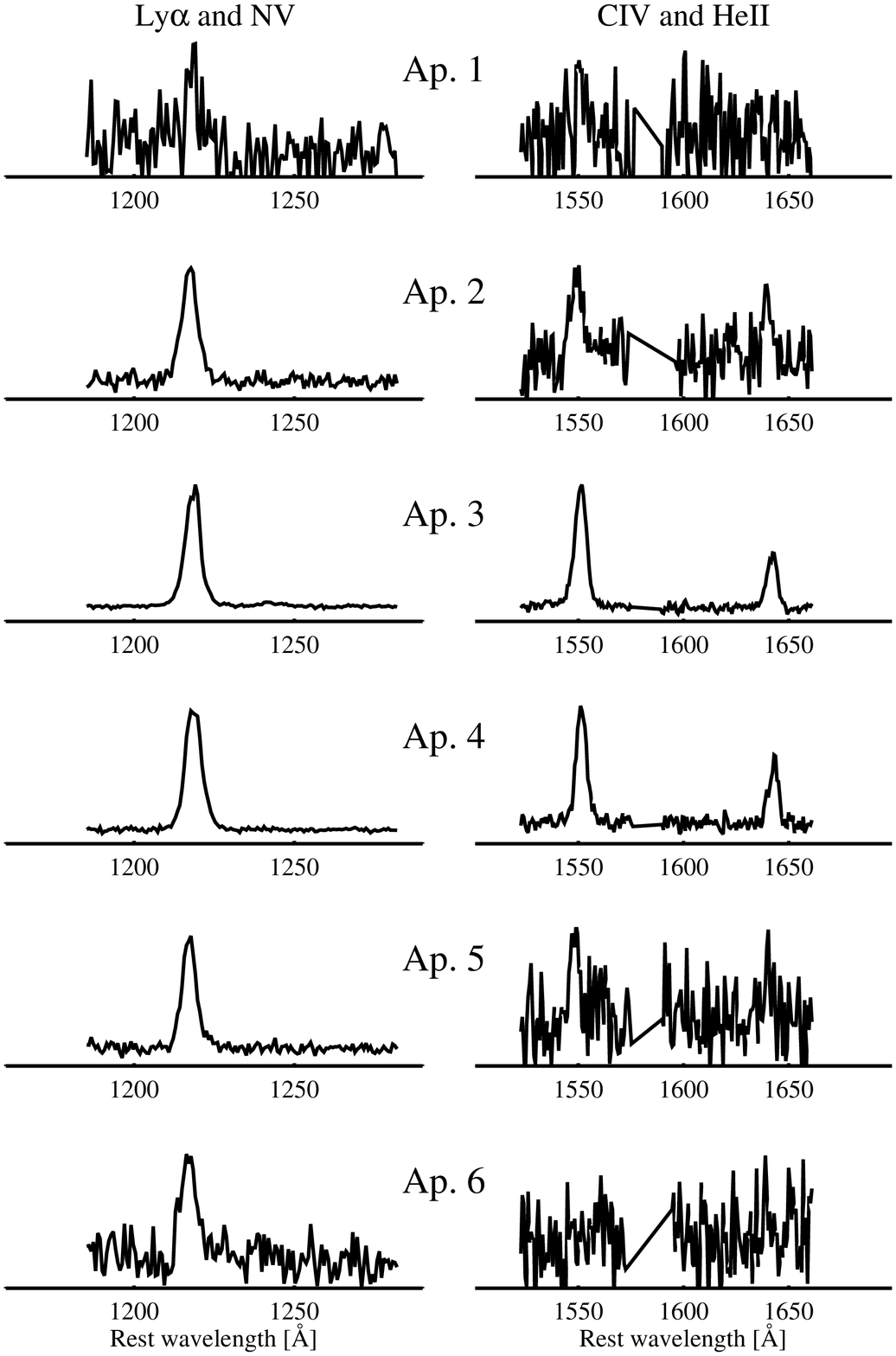}
\includegraphics{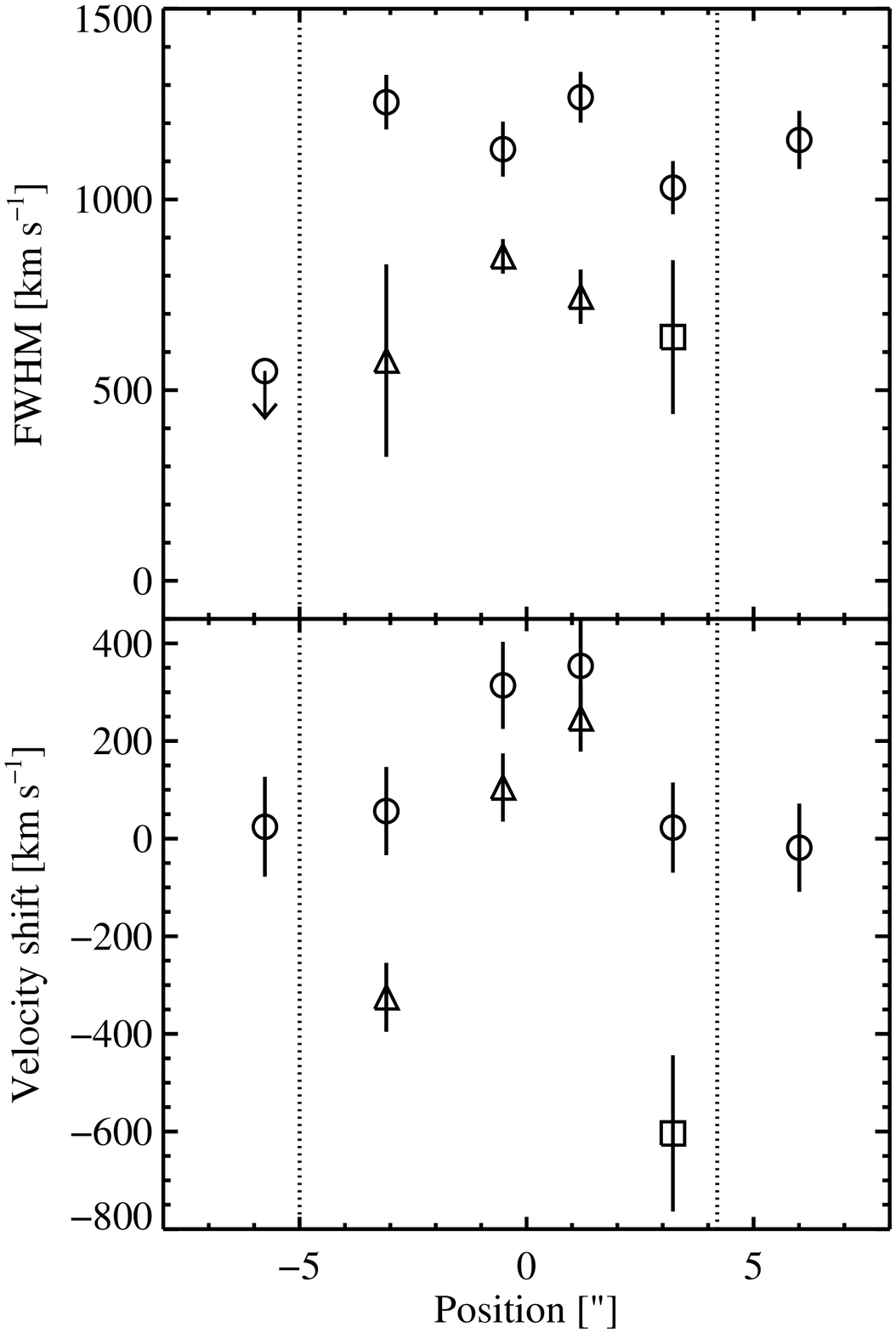}
\vspace{9in}
\caption{{\bf 1558-003:} Symbols and lines as in Fig. 1. 
Top: Notice the LSBH seen  in 
Ly$\alpha$ extending beyond the radio structures with 
apparently quieter   kinematics
compared to the high surface brightness regions. The kinematic analysis
(bottom right) shows  FWHM(HeII) in the range
550-850 km s$^{-1}$ and maximum velocity shift 550 km s$^{-1}$.
There is
some preliminary evidence for more quiescent gas in this  
object (FWHM$\leq$410 km s$^{-1}$, see text).} 
\end{figure*}

\begin{figure*}
\includegraphics{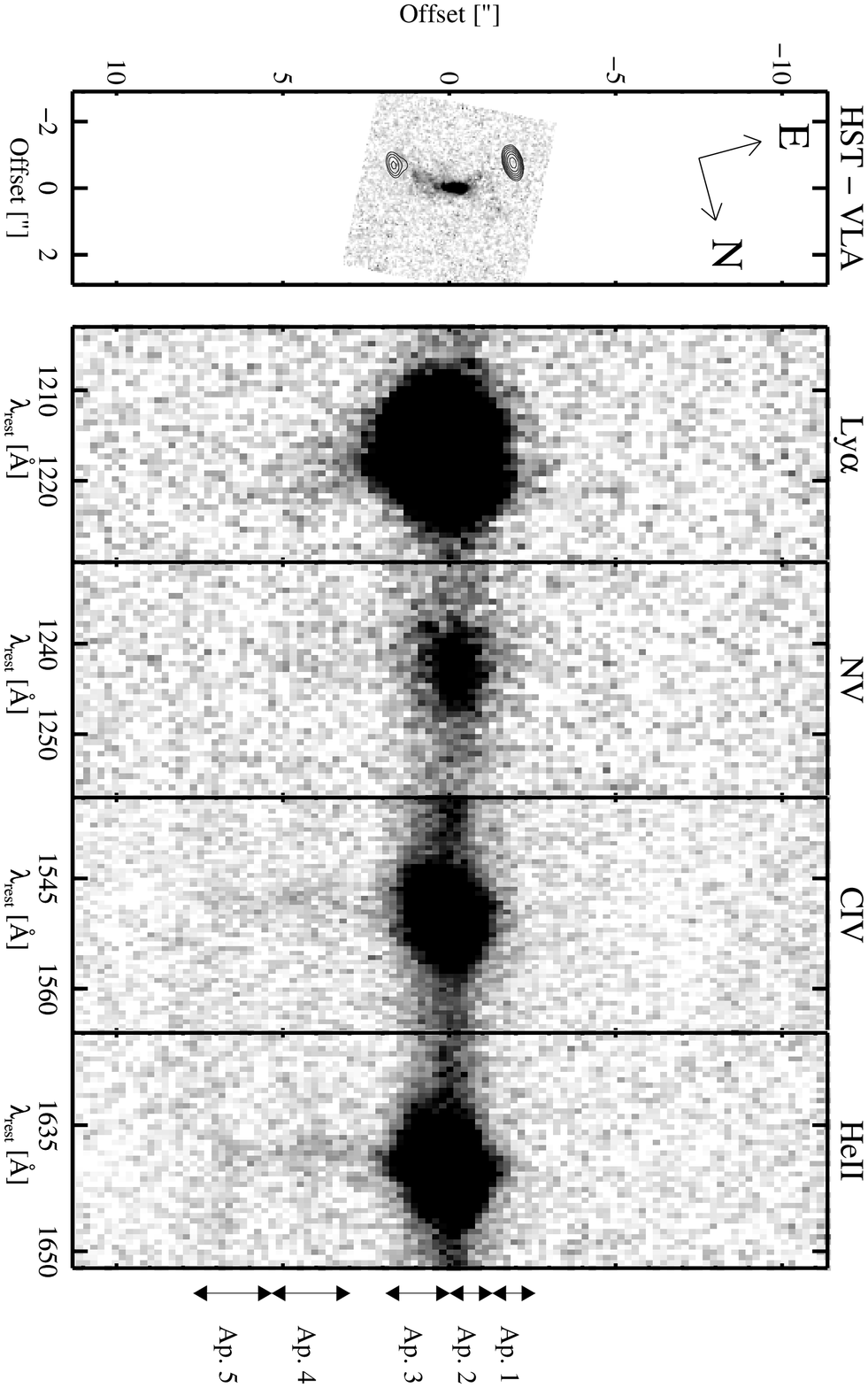}
\includegraphics{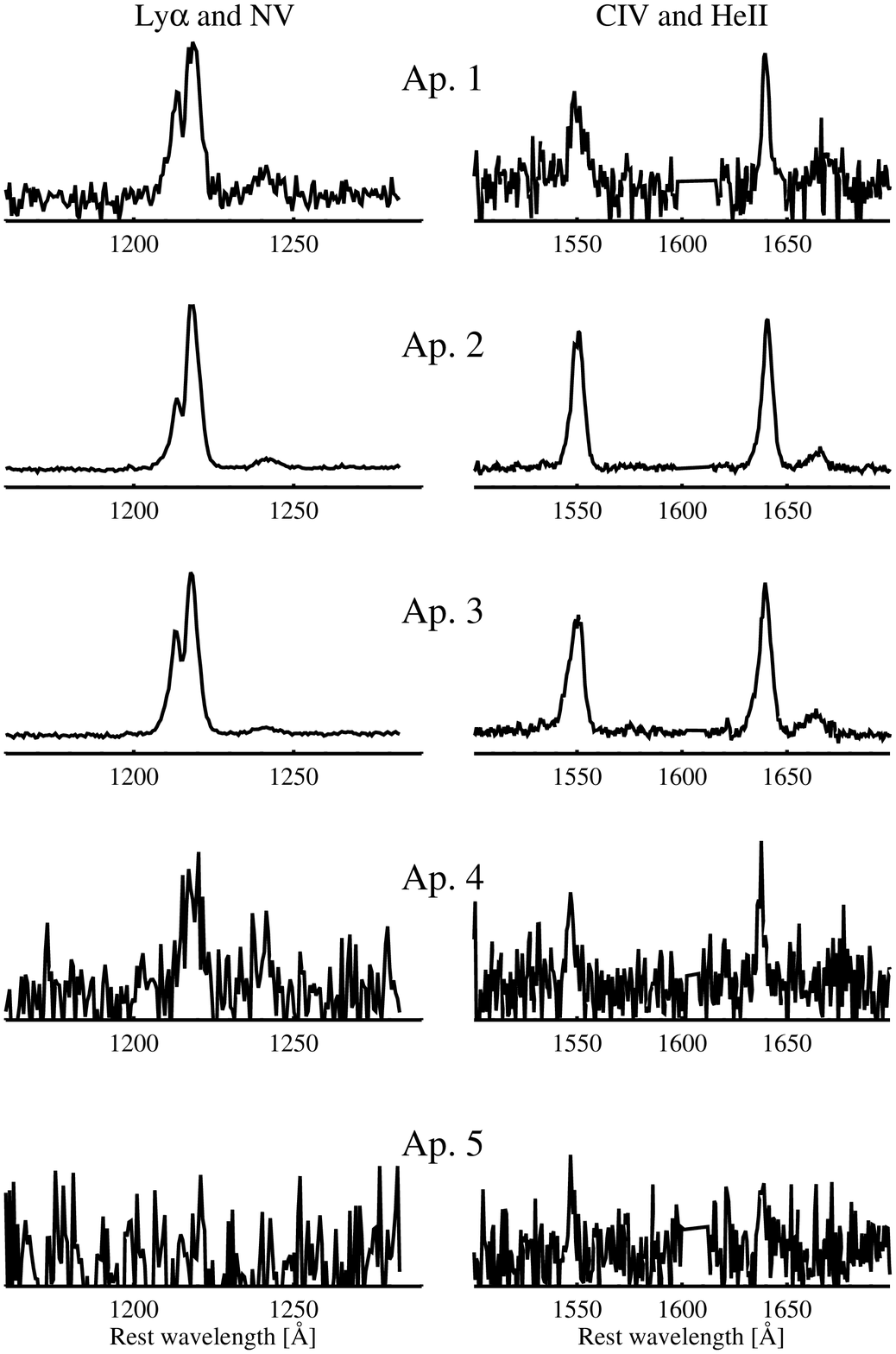}
\includegraphics{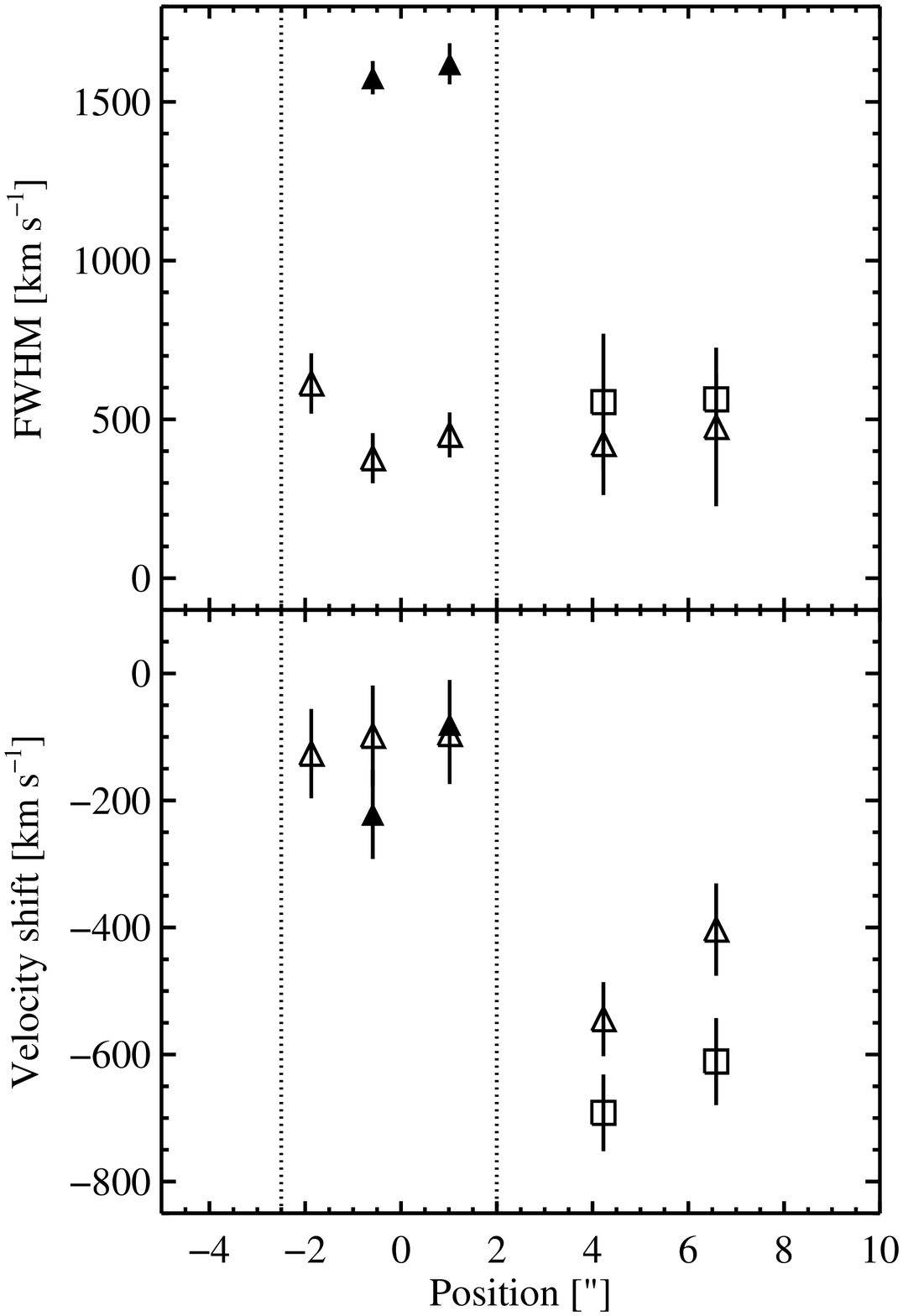}
\vspace{9in}
\caption{{\bf 0943-242:} Symbols and lines as in Fig. 1.
Notice the very extended LSBH with apparently quiescent kinematics
extending towards the west well beyond the radio structures
and in all the emission lines (top panels). A narrow component with
FWHM$\sim$400-600 km s$^{-1}$ is detected in all apertures with 
maximum velocity shift across the nebula of $\sim$450 km s$^{-1}$
(bottom right).
The results for CIV are in good agreement.
}
\end{figure*}

\begin{figure}
\includegraphics{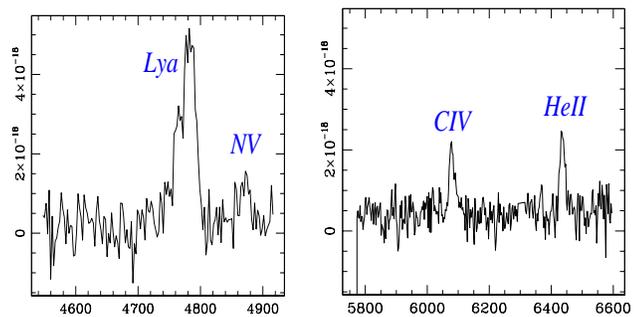}
\vspace{2in}
\caption{Spatially integrated spectrum of the quiescent LSBH in 0943-242
(emission detected beyond the radio structures).
Compared with other objects,
NV is very strong relative to CIV and HeII in this case, suggesting
high levels of chemical enrichment.}
\end{figure}
\end{document}